\documentclass[1p,number,11pt,compress]{elsarticle}
\usepackage{latexsym,amsmath,amsfonts,amssymb,graphics,color,bm,mathrsfs}
\usepackage{amsthm} 

\usepackage{tikz}

\journal{Journal of Differential Equations}

\usepackage{xcolor}              
\definecolor{a1}{rgb}{0,0,0.8}   
\usepackage{colortbl}	           
\usepackage[colorlinks=true,linktocpage=true,hyperindex=true,pageanchor=true,bookmarks=true]{hyperref} 
\hypersetup{citecolor=a1,linkcolor=a1,menucolor=a1,urlcolor=a1,runcolor=a1}
\hypersetup{citebordercolor=a1,linkbordercolor=a1,urlbordercolor=a1,runbordercolor=a1}
\hypersetup{hypertexnames=false} 

\usepackage{upgreek}
\usepackage{nicefrac}
\usepackage{tikz}
\usepackage{graphicx}
\usepackage{amsmath}
\usepackage{relsize}
\DeclareMathAlphabet{\mathpzc}{OT1}{pzc}{m}{it}
\usepackage{upgreek}
\usepackage{nicefrac}

\newtheorem{theorem}{\bf Theorem}[section]

\newtheorem{lem}{\bf Lemma}[section]

\newcommand{\be}{\begin{equation}}
\newcommand{\ee}{\end{equation}}

\DeclareMathAlphabet{\mathpzc}{OT1}{pzc}{m}{it}
\usepackage{upgreek}
\usepackage{nicefrac}

\begin{document}
\begin{frontmatter}
\title{Chemical Principle and PDE of Variational Electrodynamics}
\author[fsc]{Jayme De Luca}
\ead{jayme.deluca@gmail.com}
\address[fsc]{Departamento de F\'{\i}sica,
Universidade Federal de S\~{a}o Carlos,
S\~{a}o Carlos, S\~{a}o Paulo 13565-905, Brazil}

\begin{abstract}
 The two-body problem of variational electrodynamics possesses differential-delay equations of motion with state-dependent delays of neutral type and solutions that can have velocity discontinuities on countable sets. From a periodic orbit possessing some mild properties at breaking points, we define a synchronization function in $\mathbb{R} \times \mathbb{R}^3$, which is further used to construct two bounded oscillatory functions vanishing at breaking points and whose first derivatives are continuous and defined everywhere but at breaking points. The oscillatory functions are associated with a PDE identity in $ \mathbb{W}^{\mathpzc{2},\mathpzc{2}}( \mathbb{R}^3)$, and we postulate ordering conditions for the PDE identity to define a Fredholm-Schroedinger operator in $ \mathbb{W}^{\mathpzc{2},\mathpzc{2}}( \mathbb{R}^3)$, with a spin-orbit forcing term belonging to $\{  \mathbb{L}^2(\mathbb{R}^3) \cap O(\frac{1}{r^2}) \} $. As an application, we introduce the Chemical Principle criterion to select orbits with asymptotically vanishing far-fields and estimate the Bohr radius parameter of the Fredholm-Schroedinger PDE using the boundary-layer of orbits chosen according to the Chemical Principle criterion. Last, working backward, we derive a Chemical Principle-like condition \textit{from} the ordering conditions.
 \end{abstract}



\end{frontmatter}

\section{Introduction}
\subsection{Significance of the problem}
\label{significance}
Schroedinger's equation for the hydrogen atom \cite{Schroe} is a partial differential equation (PDE) defined on an \textit{infinite-dimensional space}. On the other hand, physics derives this PDE  from a mechanical problem with the Coulomb force, which is an ordinary differential equation (ODE) on a \textit{finite-dimensional} space. Here we derive a PDE for the electromagnetic two-body problem of variational electrodynamics \cite{JDE1,Mehra,FeyNobel} starting from a physically sensible \textit{infinite-dimensional problem}. To name a few, some reasons our approach had to wait so long were:
\begin{itemize} 
\item By the 1930s there was still no equation of motion for electrodynamics, and the equations derived later \cite{JDE1,WheeFey,JMP2009,cinderela} are infinite-dimensional problems in the case of two-body motion. 
\item The critical point conditions for the variational electromagnetic two-body problem \cite{JDE1, cinderela} involve \textit{four} state-dependent delays of neutral type, \textit{four} velocity-dependent denominators\cite{WheeFey,cinderela} and a different partial Lagrangian\cite{JDE1} for each particle. It is, therefore, a completely different problem as compared to the Coulomb-mechanical ODE problem of celestial mechanics.
\item Only after 1962 the no-interaction theorem \cite{no-interaction} exposed the severe limitations of the ODE quantization program.
\item Only after the 1970s the differential-delay equations with state-dependent delays started to be understood as infinite-dimensional problems \cite{JDE1,Mallet-Paret1,Mallet-Paret_bound-layer,JackHale,Nicola2,Dirk} (for an extensive list of references see also \cite{BellenZennaro}). 
\end{itemize}

\subsection{What is this paper about and its two main parts}
\label{parts}
\begin{enumerate}
\item  The light-cone functions of the electromagnetic two-body problem\cite{cinderela} are extended to global functions of the domain $\mathbb{R} \times \mathbb{R}^3$. Some mild properties at breaking points are sufficient to prove the existence of two synchronization functions in $\mathbb{R} \times \mathbb{R}^3$ for periodic orbits possessing a countable set of velocity discontinuities \cite{JDE1, cinderela, minimizer}. The synchronization functions are used to construct two continuous oscillatory functions that vanish at breaking points, possess continuous first derivatives everywhere but at breaking points, and further satisfy a natural PDE that includes spin-orbit terms. In order to apply the Chemical Principle criterion, the natural PDE is associated with a complex PDE identity on a Hilbert space where only the second derivatives are discontinuous, i.e., $ \mathbb{W}^{\mathpzc{2},\mathpzc{2}}( \mathbb{R}^3)$.  An asymptotic ordering is postulated, which introduces a linear approximation for the complex PDE identity and defines a Fredholm-Schroedinger linear PDE problem in $ \mathbb{W}^{\mathpzc{2},\mathpzc{2}}(\mathbb{R}^3)$. 
\item As an application of the first part, we introduce the Chemical Principle criterion to select orbits with vanishing far-fields and discuss its predictions for the spectral magnitudes of atomic physics. The interesting questions to keep in mind here are (i) the linearizing consequences of the asymptotic ordering for the PDE identity,  
(ii) \textit{how} the ordering conditions that lead to the Fredholm-Schroedinger PDE problem are compatible with orbital properties of physical interest, and (iii) how the parameters of the Fredholm-Schroedinger PDE relate to the boundary-layer of the orbit. Last, we work backward \textit{from} the ordering conditions of the Fredholm-Schroedinger PDE and derive a generalized Chemical Principle condition, thus relating the two parts of the paper. \end{enumerate}

\subsection{Chemical Principle, Schroedinger equation and spectroscopic lines}
\label{chemical+Schroe+spectro}
 As we review in Appendix \ref{Erdman}, the Li\'{e}nard-Wiechert vector fields of a point charge are functions $ \vec{V}: (t,\mathbf{x}) \in \mathbb{R}^4 \rightarrow \mathbb{R}^3$ which split in a transversal vector field with modulus decreasing as $1/r_{\mbox{\tiny $cone$}}(t,\mathbf{x})$, henceforth the \textit{far-field}, plus a reminder field with modulus decreasing as $1/r_{\mbox{\tiny$cone$}}^2(t,\mathbf{x})$,  henceforth the \textit{near-field}, where $r_{\mbox{\tiny$cone$}}(t,\mathbf{x}): \mathbb{R}^4 \rightarrow \mathbb{R}$ is the distance in light-cone from the point charge\cite{Jackson}. An example of a bounded two-body orbit possessing far-fields that do not vanish asymptotically is the $C^\infty$ circular orbit \cite{Schild,Hans}. 
 \par
 The Chemical Principle criterion quantifies the influence of a two-body orbit on the other charges of the universe. Its definition starts from the restricted three-body problem by placing a (test) third charge to suffer the far-fields of the original two charges along the given two-body orbit \textit{without} creating fields to act back on the original two charges. Variational electrodynamics would describe the full three-body problem \textit{if} one added a third charge to the two-body problem\cite{cinderela} \textit{and} let it cause fields on the original two charges. The Chemical Principle criterion is thus motivated by the intermediate $2\nicefrac{1}{2}$-body infinite-dimensional problem.
 \par
  Because of the importance of the Schroedinger equation in physics and as a hard test for a condition of the Chemical Principle type, in \S \ref{applications_Chemical_principle}-\ref{tested_consequences} we give a crash review of how some resonances along orbits with vanishing far-fields predict the magnitudes of the spectral lines of hydrogen\cite{cinderela}.

 \subsection{How the paper is divided}
In \S \ref{Basic}-\ref{light-cone_section} we introduce the light-cone condition and associated delay functions, prove theorem \ref{lema1} about the existence and uniqueness of the delay functions. We also prove theorem \ref{lemmazero} on the zeros of the delay functions and theorem \ref{Fi_lower_bound} on a lower bound for the delay functions. In \S \ref{Basic}-\ref{derivatives_of_delay_functions} we calculate the time derivative, the first and second spatial derivatives, and the Laplacian derivative of the delay functions for continuous and piecewise differentiable trajectories. In \S \ref{Basic}-\ref{musical_properties} we introduce the properties necessary for distributional synchronization when the velocities are discontinuous. In \S \ref{Basic}-\ref{syncFu} we introduce the synchronization function and prove the musical Lemma \ref{musicallema} on the synchronization of breaking points. We also prove that the natural oscillatory functions and their gradients and time derivatives are continuous and bounded, which is the material of theorem \ref{CONTI_BOUNDS}. In theorem \ref{mystery_theorem} we obtain a relation between space derivatives and time derivatives, and theorem \ref{Sobolema} shows that the distributions associated with the musical orbit belong to $ \mathbb{W}^{\mathpzc{2},\mathpzc{2}}(\mathcal{B})$, where $\mathcal{B}$ is a ball containing the protonic trajectory and the image trajectory. In \S \ref{PDE} we discuss the natural PDEs of the musical orbit. In \S \ref{PDE}-\ref{natural_PDE} we formulate two natural PDEs in $ \mathbb{W}^{\mathpzc{2},\mathpzc{2}}(\mathcal{B})$, whose left-hand sides are suited for a linear approximation. In \S \ref{PDE}-\ref{estima_e_classifica} we discuss a qualitative classification of the reminders and introduce the linear bridging approximation. In \S  \ref{PDE}-\ref{linear_PDE}  we extend the natural PDEs from $ \mathbb{W}^{\mathpzc{2},\mathpzc{2}}(\mathcal{B})$ to $\mathbb{W}^{\mathpzc{2},\mathpzc{2}}(\mathbb{R}^3)$, which extension is used in \S \ref{PDE}-\ref{identity} to derive a complex PDE identity. In \S \ref{PDE}-\ref{Fredholm-Schroe} we postulate ordering conditions for the complex PDE identity, which define a Fredholm-Schroedinger problem with an orbit-dependent forcing term. In \S \ref{applications_Chemical_principle} we discuss applications of the Fredholm-Schroedinger PDE problem. In \S \ref{applications_Chemical_principle}-\ref{definition_of_Chemical} we introduce the Chemical Principle criterion to select orbits with vanishing far-fields and in \S \ref{applications_Chemical_principle}-\ref{quasi-semiflow} we derive a condition equivalent to the Chemical Principle condition, here called the quasi-semiflow condition. In \S \ref{applications_Chemical_principle}-\ref{tested_consequences} we discuss some tested consequences of the Chemical Principle criterion on the prediction of spectroscopic lines with the boundary-layer perturbation theory of Ref. \cite{cinderela}. In \S \ref{applications_Chemical_principle}-\ref{generalized} we work in the opposite direction and derive a generalized Chemical Principle condition \textit{from} the ordering conditions of the Fredholm-Schroedinger PDE problem. In \S \ref{applications_Chemical_principle}-\ref{linearization_estimate}
we exhibit the linearizing consequences of the ordering conditions to the natural PDE and use the ordering conditions to define the Bohr radius parameter from the boundary-layer of the musical orbit. We exhibit a rough estimate for the Bohr radius parameter obtained by approximating the musical orbits with some orbits with boundary layers obtained in Ref. \cite{cinderela}.  In \S \ref{Sumasection}-\ref{sumario} we summarize our results and in \S \ref{Sumasection}-\ref{discussions} we put the discussions, conjectures and conclusion. In Appendix \ref{asymptotics} we derive several identities and asymptotic expansions of delay functions which are used in the paper.  In Appendix \ref{Erdman} we review the Wheeler-Feynman equations of motion, the Weierstrass-Erdmann corner conditions, the Li\'{e}nard-Wiechert vector fields of a point charge, and the many-body problem of variational electrodynamics. 
\section{Delay functions and distributional synchronization} 
\label{Basic}
\subsection{The light-cone condition and associated delay functions}
\label{light-cone_section}

We use a unit system where the speed of light is $c \equiv 1$ and the electronic charge and mass are $e_\mathpzc{1} \equiv-1$ and $m_{\mathpzc{1}} \equiv 1$ while the protonic charge and mass are $e_\mathpzc{2}=1$ and $m_\mathpzc{2} = M_\mathpzc{p} $, respectively. The accepted value for the protonic mass is about $M_\mathpzc{p} \simeq 1837$ in our unit system but our calculations are made with an arbitrary protonic mass. We work on the space $\mathbb{R} \times \mathbb{R}^3$ where every point has a time $t$ obtained by Einstein synchronization of clocks and spatial coordinates $\mathbf{x} \in \mathbb{R}^3 \equiv (x,y,z)  $. Henceforth a bounded trajectory is a continuous function $\mathbf{x}_\mathpzc{j} (t) : \mathbb{R} \rightarrow \mathbb{R}^3$ with a bounded image. We reserve the name \textit{orbit} for a set made of one trajectory for each particle further satisfying the Wheeler-Feynman equations of motion and the Weierstrass-Erdmann corner conditions reviewed in Appendix \ref{Erdman}. The sub-index $\mathpzc{j}$ is henceforth used to distinguish the particles and we shall use equivalently either $\mathpzc{j}=\mathpzc{1}$ or $\mathpzc{j}=\mathpzc{e}$ to denote the electronic quantities and either $\mathpzc{j}=\mathpzc{2}$ or $\mathpzc{j}=\mathpzc{p}$ to denote the protonic quantities. The index is sometimes extended to $\mathpzc{j}=\upmu$ in order to denote the image trajectory.
\par
State-dependent delays appear in variational electrodynamics because of the light-cone condition. Even though the light-cone construction has a meaning for any chase problem, electrodynamics is sensible only for bounded trajectories possessing a velocity defined almost everywhere by a number smaller than the speed of light $c \equiv 1$, i.e.,
{\small
\begin{eqnarray}
\vert \dot{\mathbf{x}_\mathpzc{j}}(t) \vert < 1,
\label{subluminal_cond}
\end{eqnarray}}
\noindent
for $\mathpzc{j} \in \{ \mathpzc{p}, \mathpzc{e}  \}$ with the possible exception of a countable set, henceforth a sub-luminal trajectory.  For each sub-luminal trajectory $\mathbf{x}_\mathpzc{j}(t)$ we can define the following two light-cone conditions: ($\mathpzc{1}$) the \textit{advanced} light-cone condition is the time for a light signal emitted at point $(t,\mathbf{x})$ to intersect the trajectory $\mathbf{x}_\mathpzc{j}(t)$ at the later time $t_\mathpzc{j}^+(t, \mathbf{x}) \equiv t+ \phi_\mathpzc{j}^{+}(t,\mathbf{x})$ and ($\mathpzc{2}$) the \textit{retarded} light-cone condition is the time for a signal emitted by the trajectory at an earlier time $t_\mathpzc{j}^-(t,\mathbf{x}) \equiv t-\phi_\mathpzc{j}^{-}(t,\mathbf{x}) $ to arrive at point $(t,\mathbf{x})$. The former deviating arguments define light-cone maps $ t_\mathpzc{j}^\pm(t,\mathbf{x}) : \mathbb{R} \times \mathbb{R}^3 \rightarrow \mathbb{R}$ by
{\small
\begin{equation}
t_\mathpzc{j}^\pm(t,\mathbf{x}) \equiv t \pm \phi_\mathpzc{j}^{\pm}(t,\mathbf{x}),
\label{light-cone_map}
\end{equation}}
where the \textit{delay functions} $\phi_\mathpzc{j}^{\pm}(t,\mathbf{x})$ are defined implicitly by
{\small
\begin{eqnarray}
\phi_\mathpzc{j} ^{\pm} (t,\mathbf{x}) \equiv  \vert {{\mathbf{x} -\mathbf{x}}_\mathpzc{j}(t_\mathpzc{j}^\pm)} \vert = | t-t_\mathpzc{j}^\pm  | ,
\label{defi}
\end{eqnarray}}
\noindent
as the Euclidean norm of the spatial separation in $\mathbb{R}^3$, thus defining continuous functions $ \phi_\mathpzc{j}^{\pm}(t,\mathbf{x}) : \mathbb{R} \times \mathbb{R}^3 \rightarrow \mathbb{R}$.
Here we study \textit{bounded} continuous and piecewise-differentiable sub-luminal trajectories possessing velocity discontinuities on a countable set of points. For these, one solution to either light-cone condition (\ref{light-cone_map}) is the accumulation point of the bounded iterative series $\{ t_{n}^{\pm} \} $ defined by $t_{1}^{\pm} \equiv t $ for $n=1$ and recursively for $n>1$ by $t_{n+1}^{\pm} \equiv t  \pm \vert \mathbf{x}-{\mathbf{x}}_\mathpzc{j}(t_{n}^{\pm}) \vert  $. In the following we prove that the above iterative solutions of (\ref{light-cone_map}) are unique along sub-luminal orbits.

\begin{theorem} For a bounded sub-luminal trajectory $\mathbf{x}_\mathpzc{j}(t)$, the advanced and the retarded light-cone conditions (\ref{defi}) have unique solutions $t_\mathpzc{j}^\pm(t, \mathbf{x})$. 
\label{lema1}
\end{theorem}
\begin{proof} We start with the advanced light-cone of a given bounded sub-luminal trajectory $\mathbf{x}_\mathpzc{j}(s)$, defined by (\ref{defi}) and (\ref{light-cone_map}) with the plus sign. One solution exists and is given by the iterative solution described above.  Our proof by contradiction assumes that there are two different solutions to the advanced light-cone condition defined by Eq. (\ref{light-cone_map}) with the plus sign at a given $(t,\mathbf{x})$, i.e.,   $t_a= t+ \vert \mathbf{x}-{\mathbf{x}}_\mathpzc{j}(t_a) \vert$  and $t_b=t+ \vert \mathbf{x}-{\mathbf{x}}_\mathpzc{j}(t_b) \vert$. We have the inequalities 
{\small
\begin{eqnarray}
t_b-t_a = \vert \mathbf{x}-{\mathbf{x}}_\mathpzc{j}(t_b) \vert -\vert \mathbf{x}-{\mathbf{x}}_\mathpzc{j}(t_a) \vert \leq \vert   \mathbf{x}_\mathpzc{j}(t_b)-{\mathbf{x}}_\mathpzc{j}(t_a)    \vert \notag \\ \leq \vert \int_{t_a}^{t_b} \dot{\mathbf{x}}_\mathpzc{j}(s)ds \vert  \leq  \int_{t_a}^{t_b} \vert \dot{\mathbf{x}}_\mathpzc{j}(s)\vert ds  \leq t_b-t_a. \label{qtp}
\end{eqnarray}}
The inequality sign holds on the right-hand side of (\ref{qtp}) when $t_b-t_a \neq 0$ because the velocity satisfies (\ref{subluminal_cond}) almost everywhere. In view of the left-hand side of (\ref{qtp}), the only sensible alternative is $t_b-t_a=0$  and thus the solution must be unique. The proof of uniqueness of solutions of (\ref{light-cone_map}) for the retarded light-cone is analogous. 
\end{proof}

\begin{theorem}The conditions $\phi_\mathpzc{j}^{\pm}(t,\mathbf{x})=0$ obtained from (\ref{defi}) have the unique solution $\mathbf{x}=\mathbf{x}_\mathpzc{j}(t)$, and imply that $t_\mathpzc{j}^\pm(t,\mathbf{x})=t$.
\label{lemmazero}
\end{theorem}
\begin{proof}
The functions $t_\mathpzc{j}^\pm(t,\mathbf{x})=t\pm\phi_\mathpzc{j}^{\pm}(t,\mathbf{x})$ and $  \phi_\mathpzc{j}^{\pm}(t,\mathbf{x}) \geq 0$ defined respectively by Eqs. (\ref{light-cone_map}) and (\ref{defi}) are continuous because the trajectories are continuous. For either the plus or the minus sign cases, the condition $\phi_\mathpzc{j}^{\pm}(t,\mathbf{x})=|\mathbf{x}-\mathbf{x}_\mathpzc{j}(t)|= 0$ has the unique solution $\mathbf{x}=\mathbf{x}_\mathpzc{j}(t)$, which implies $t_\mathpzc{j}^\pm(t,\mathbf{x})=t$. 
\end{proof}
The following lower bound for the $\phi_\mathpzc{j}^{\pm}(t,\mathbf{x})$ is useful. 
\begin{theorem} A lower-bound for the delay functions defined by (\ref{defi}) is  given by $\phi_\mathpzc{j}^{\pm}(t,\mathbf{x}) \geq  |\mathbf{x}- \mathbf{x}_\mathpzc{j}(t)| /2$.
\label{Fi_lower_bound}
\end{theorem}
\begin{proof} Starting from (\ref{defi}) and using the triangular inequality we have 
{\small
\begin{eqnarray}
\phi^\pm_\mathpzc{j}(t,\mathbf{x}) =|\mathbf{x}-\mathbf{x}_\mathpzc{j}(t_\mathpzc{j} \pm \phi^\pm_\mathpzc{j}) | \geq |\mathbf{x}-\mathbf{x}_\mathpzc{j}(t)|-|\mathbf{x}_\mathpzc{j}(t)-\mathbf{x}_\mathpzc{j}(t \pm \phi^\pm_\mathpzc{j})| \notag \\ 
 \geq |\mathbf{x}-\mathbf{x}_\mathpzc{j}(t)| -\phi^\pm_\mathpzc{j}(t,\mathbf{x}), \label{linha5}
\end{eqnarray}}
where the last inequality holds when $\mathbf{x}_\mathpzc{j}(t)$ possesses a sub-luminal velocity defined almost everywhere inside either $[t, t + \phi_\mathpzc{j}^+]$ or $[t-\phi_\mathpzc{j}^-, t]$. The proof is finalized by passing the last $\phi^\pm_\mathpzc{j}(t,\mathbf{x})$ to the left-hand side of (\ref{linha5}) and dividing by two for either the plus-sign case or the minus-sign case. 
\end{proof}
A corollary of theorem \ref{Fi_lower_bound} is the upper bound 
{\small
\begin{eqnarray}
\frac{1}{\phi^\pm_\mathpzc{j}(t,\mathbf{x}) } < \frac{2}{|\mathbf{x}-\mathbf{x}_\mathpzc{j} (t) |}. \label{COROBOUNDS}
\end{eqnarray}}

\subsection{Derivatives of the delay functions}
\label{derivatives_of_delay_functions}
  Starting from the implicit definition (\ref{defi}) and assuming the bounded trajectory $\mathbf{x}_\mathpzc{j} (t)$ possesses a velocity at time $t_\mathpzc{j}^\pm(t,\mathbf{x}) =t \pm \phi_\mathpzc{j}^{\pm}(t,\mathbf{x})$,  the gradient and time-derivative of $\phi_\mathpzc{j} ^{\pm}(t,\mathbf{x})$ evaluate to
  {\small
 \begin{eqnarray}
\vec{\nabla} \phi_\mathpzc{j}^{\pm}(t,\mathbf{x})&=&\frac{\hat{\mathbf{n}}_\mathpzc{j}^{\pm}}{(1 \pm \hat{\mathbf{n}}_\mathpzc{j}^\pm \cdot \mathbf{v}_\mathpzc{j}^\pm)},
\label{gradient} \\
\frac{ \partial \phi_\mathpzc{j}^{\pm}}{\partial t}(t,\mathbf{x})&=&-\frac{\hat{\mathbf{n}}_\mathpzc{j}^\pm \cdot \mathbf{v}_\mathpzc{j}^\pm}{(1 \pm \hat{\mathbf{n}}_\mathpzc{j}^\pm \cdot \mathbf{v}_\mathpzc{j \pm})}=\pm \left( |\vec{\nabla} \phi_\mathpzc{j}^\pm| -1\right),
\label{dfidt}
\end{eqnarray}
}
where {\small $\mathbf{v}^\pm_\mathpzc{j} \equiv \frac{d\mathbf{x}_\mathpzc{j}}{dt}|_{t=t^\pm_\mathpzc{j}(t,\mathbf{x})}$} and $\hat{\mathbf{n}}_\mathpzc{j}^\pm(t,\mathbf{x})$ stands for a vector field of unitary modulus defined by 
{\small
\begin{eqnarray}
\hat{\mathbf{n}}_\mathpzc{j}^{\pm}(t,\mathbf{x}) \equiv \frac{ \mathbf{x} -\mathbf{x}_\mathpzc{j}^\pm }{       |\mathbf{x} -\mathbf{x}_\mathpzc{j}^\pm |    }.
\label{nj}
\end{eqnarray}}
\par
To simplify the notation, the Cartesian $x$-component of the vector field $\hat{\mathbf{n}}_\mathpzc{j}^{\pm}(t, \mathbf{x})$ defined by Eq. (\ref{nj}) is henceforth denoted without either the over-hat or the indices $\pm$ or the indices $\mathpzc{j}$, i.e.,
{\small
\begin{eqnarray}
n_x \equiv \frac{x-x_\mathpzc{j}}{   |\mathbf{x} -\mathbf{x}_\mathpzc{j} |  },
\label{nx}
\end{eqnarray}}
with analogous definitions for the $n_y$ and $n_z$ components. 
The spatial derivatives of $n_x$ are
{\small
\begin{eqnarray}
\frac{ \partial n_x}{\partial x}& =& \frac{1}{\phi^\pm_\mathpzc{j}}-\frac{n_x^2 \pm n_x v_{jx}}{\phi^\pm_\mathpzc{j} (1 \pm \hat{\mathbf{n}}^\pm_\mathpzc{j} \cdot \mathbf{v}^\pm_\mathpzc{j})}, \label{nxx} \\
\frac{ \partial n_x}{\partial y}& =& -\frac{n_x n_y \pm n_x v_{jy}}{\phi^\pm_\mathpzc{j} (1 \pm \hat{\mathbf{n}}^\pm_\mathpzc{j} \cdot \mathbf{v}^\pm_\mathpzc{j})}, \label{nxy} \\
\frac{ \partial n_x}{\partial z}& =& -\frac{n_x n_z \pm n_x v_{jz}}{\phi^\pm_\mathpzc{j} (1 \pm \hat{\mathbf{n}}^\pm_\mathpzc{j} \cdot \mathbf{v}^\pm_\mathpzc{j})} \label{nxz}.
\end{eqnarray} }
The other six combinations of derivatives are obtained by permuting the indices $x$, $y$ and $z$ in Eqs. (\ref{nxx}), (\ref{nxy}) and (\ref{nxz}). For example, using Eq. (\ref{nxx}) and its $y$ and $z$ versions we have
{\small
\begin{eqnarray}
\vec{\nabla} \cdot \hat{\mathbf{n}}_\mathpzc{j} ^{\pm}=\frac{\partial n_x}{\partial x} + \frac{\partial n_y}{\partial y} +\frac{\partial n_z}{\partial z}=\frac{2}{\phi_\mathpzc{j}^\pm}.
 \label{divergence}
\end{eqnarray}}
At points where $\mathbf{x}_\mathpzc{j}(t_\mathpzc{j})$ possesses a derivative {\small $\mathbf{v}^\pm_\mathpzc{j} \equiv \frac{d\mathbf{x}_\mathpzc{j}}{dt}|_{t=t^\pm_\mathpzc{j}(t,\mathbf{x})}$}, the time-derivative of $\hat{\mathbf{n}}^\pm_\mathpzc{j}(t,\mathbf{x})$ can be evaluated from (\ref{nj}), yielding
{\small
\begin{eqnarray}
\frac{\partial \hat{\mathbf{n}}^\pm_\mathpzc{j}}{\partial t}=-\frac{\mathbf{v}^\pm_\mathpzc{j}}{\phi^\pm_\mathpzc{j}} +\frac{ \hat{\mathbf{n}}^\pm_\mathpzc{j} \cdot \mathbf{v}^\pm_\mathpzc{j} }{\phi^\pm_\mathpzc{j} (1 \pm \hat{\mathbf{n}}^\pm_\mathpzc{j} \cdot \mathbf{v}^\pm_\mathpzc{j})}(\hat{\mathbf{n}}^\pm_\mathpzc{j} \pm \mathbf{v}^\pm_\mathpzc{j})= \frac{    \hat{\mathbf{n}}^\pm_\mathpzc{j} \times( \hat{\mathbf{n}}^\pm_\mathpzc{j} \times \mathbf{v}^\pm_\mathpzc{j} )       }{\phi^\pm_\mathpzc{j} (1 \pm \hat{\mathbf{n}}^\pm_\mathpzc{j} \cdot \mathbf{v}^\pm_\mathpzc{j})} \label{dndt}.
\end{eqnarray}}
At points where $\mathbf{x}_\mathpzc{j}(t_\mathpzc{j})$ possesses an acceleration {\small $\mathbf{a}^\pm_\mathpzc{j} \equiv d^2 \mathbf{x}_\mathpzc{j}(t)/dt^2 |_{t=t_\mathpzc{j}^\pm(t,\mathbf{x})}$} defined at $t_\mathpzc{j}^\pm(t,\mathbf{x})$, the \textit{second} time-derivative of the $\phi^\pm_\mathpzc{j}(t,\mathbf{x})$ can be evaluated from the last term of the right-hand side of (\ref{dfidt}) using (\ref{dndt}), yielding
{\small
\begin{eqnarray}
\frac{\partial^2 \phi^\pm_\mathpzc{j} }{\partial t^2}& =&\frac{|\hat{\mathbf{n}}^\pm_\mathpzc{j} \times \mathbf{v}^\pm_\mathpzc{j} |^2 }{\phi^\pm_\mathpzc{j}(1 \pm \hat{\mathbf{n}}^\pm_\mathpzc{j} \cdot \mathbf{v}^\pm_\mathpzc{j})^3}- \frac{\hat{\mathbf{n}}^\pm_\mathpzc{j} \cdot \mathbf{a}^\pm_\mathpzc{j}}{(1 \pm \hat{\mathbf{n}}^\pm_\mathpzc{j} \cdot \mathbf{v}^\pm_\mathpzc{j})^3}, \label{d2fidt2}
\end{eqnarray}}
\noindent and the second derivatives $\partial^2 _x \phi^\pm_\mathpzc{j}$ and $\partial^2_{xt} \phi^\pm_\mathpzc{j}$ can be evaluated from the $x$-component of (\ref{gradient}), yielding
{\small
\begin{eqnarray}
\frac{\partial^2 \phi^\pm_\mathpzc{j}}{\partial x^2}&=&\left(\frac{1}{1\pm \hat{\mathbf{n}}^\pm_\mathpzc{j} \cdot \mathbf{v}^\pm_\mathpzc{j}  }\right)\frac{\partial n_x}{\partial x} \pm n_x \frac{\partial^2 \phi^\pm_\mathpzc{j}}{\partial x \partial t}, \label{d2fidxx}\\
\frac{\partial^2 \phi^\pm_\mathpzc{j}}{\partial x \partial t}&=& \left( \frac{1}{1\pm \hat{\mathbf{n}}^\pm_\mathpzc{j} \cdot \mathbf{v}^\pm_\mathpzc{j}  } \right)  \frac{\partial n_x}{\partial t} \pm n_x \frac{\partial^2 \phi^\pm_\mathpzc{j}}{ \partial t^2} \label{d2fidxt}.
\end{eqnarray}
}
 Again for orbits possessing two piecewise-defined derivatives, the mixed spatial derivatives of $\phi^\pm_\mathpzc{j}(t,\mathbf{x})$ are obtained starting from Eqs. (\ref{nxx}), (\ref{nxy}) and (\ref{nxz}) to calculate the gradient of the scalar products $\hat{\mathbf{n}}^\pm_\mathpzc{j} \cdot \mathbf{v}^\pm_\mathpzc{j}$, yielding
 {\small
 \begin{eqnarray}
 \vec{\nabla} (\hat{\mathbf{n}}^\pm_\mathpzc{j} \cdot \mathbf{v}^\pm_\mathpzc{j}) =\frac{\mathbf{v}^\pm_\mathpzc{j}}{\phi^\pm_\mathpzc{j}} \pm \left( \frac{\hat{\mathbf{n}}^\pm_\mathpzc{j} \cdot \mathbf{a}^\pm_\mathpzc{j}}{(1\pm\hat{\mathbf{n}}^\pm_\mathpzc{j} \cdot \mathbf{v}^\pm_\mathpzc{j})} - \frac{ (|\mathbf{v}^\pm_\mathpzc{j}|^2 \pm \hat{\mathbf{n}}^\pm_\mathpzc{j} \cdot \mathbf{v}^\pm_\mathpzc{j})}{\phi^\pm_\mathpzc{j} (1\pm\hat{\mathbf{n}}^\pm_\mathpzc{j} \cdot \mathbf{v}^\pm_\mathpzc{j}) } \right) \hat{\mathbf{n}}^\pm_\mathpzc{j}  \label{gradnv},
 \end{eqnarray}
 }
 which can be used together with (\ref{gradient}) to evaluate the partial second derivative $\partial^2_{xy} \phi^\pm_\mathpzc{j}=\partial^2_{yx}\phi^\pm_\mathpzc{j}$ with the symmetric formula 
 {\small
 \begin{eqnarray}
 \frac{\partial ^2 \phi^\pm_\mathpzc{j}}{\partial y \partial x}=-\left( \frac{\hat{\mathbf{n}}^\pm_\mathpzc{j} \cdot \mathbf{a}^\pm_\mathpzc{j}}{(1\pm\hat{\mathbf{n}}^\pm_\mathpzc{j} \cdot \mathbf{v}^\pm_\mathpzc{j})^3}\right) n_x n_y
 +\frac{1}{\phi^\pm_\mathpzc{j} (1\pm\hat{\mathbf{n}}^\pm_\mathpzc{j} \cdot \mathbf{v}^\pm_\mathpzc{j})^2} \left( \frac{n_x n_y |\mathbf{v}^\pm_\mathpzc{j}|^2 }{(1\pm\hat{\mathbf{n}}^\pm_\mathpzc{j} \cdot \mathbf{v}^\pm_\mathpzc{j})}  \mp (n_x v_y+n_y v_x) \right).\notag \\  \label{mixed_deriv}
 \end{eqnarray}
 }
 Some additional relations involving the delay functions and their first and second derivatives are obtained in Appendix \ref{asymptotics}.
 
\subsection{Orbital properties for distributional synchronization}
 \label{musical_properties}
 
   Orbits of variational electrodynamics can involve {\it{absolutely continuous}} trajectories possessing velocity discontinuities of bounded variation and Lebesgue-integrable accelerations \cite{JDE1}. Here we consider only orbits whose trajectories possess two derivatives defined piecewise. The delay functions $\phi_\mathpzc{j}(t,\mathbf{x})$ must be defined as distributions in order to accommodate the following three situations: (a) the velocity is \textit{discontinuous} on a countable set and thus only the distributional version of the gradient (\ref{gradient}) is sensible, (b) when the velocity reaches the speed of light, the gradient (\ref{gradient}) has a divergent denominator and (c) the second derivatives of $\phi_\mathpzc{j}$ include singularities of type $\frac{1}{r}$, which belong to $\mathbb{L}^2_{loc}(\mathbb{R}^3)$ as found by substituting (\ref{nxx}) into (\ref{d2fidxx}) and also in formula (\ref{mixed_deriv}).  We distinguish three classes of periodic orbits: (i) the velocities are continuous and have a modulus smaller than the speed of light everywhere, henceforth type-(i) orbits, (ii) the velocities are lesser than the speed of light wherever defined and discontinuous on a countable set of breaking points, henceforth type-(ii) orbits and (iii) the velocities are continuous everywhere \textit{and} both particles reach the speed of light at a central collision point, henceforth \textit{regular collisional} or type-(iii) orbits, e.g. the orbits studied numerically in \cite{EFY2}.  
  \par
  The presence of velocity discontinuities is an obstacle for the existence of a regular distribution in $ \mathbb{W}^{\mathpzc{2},\mathpzc{2}}(\mathcal{B})$ because of a delta-function distribution coming from the piecewise integration by parts necessary to re-arrange the distributional derivative. For type-(ii) orbits we need some properties in order to construct a regular distribution. These are henceforth called \textit{musical properties}, i.e.,
 \begin{enumerate}
 \item[$(P1)$]  Both trajectories are periodic and have the same integer number $2N_{\flat}$ of breaking points inside each period $T$.
\item[$(P2)$] There exists a $\upmu \in \mathbb{R}$ and a bounded $C^2$ vector function $\mathtt{G}_\mathsmaller{B}(t)$, henceforth the \textit{guiding} trajectory, and a globally defined \textit{image} trajectory 
{\small
 \begin{eqnarray}
 \mathbf{X}_\upmu(t) &\equiv&  -{\mathtt{G}}_\mathsmaller{B}(t) + \upmu  \mathbf{x}_\mathpzc{e}(t) ,  \, \, \forall \, t \in \mathbb{R}, \label{defimageXmu}
  \end{eqnarray}}
which is sub-luminal, i.e., 
 {\small
 \begin{eqnarray}
 |d\mathbf{X}_\upmu/dt| =|\dot{\mathtt{G}}_\mathsmaller{B} -\upmu  \mathbf{v}_\mathpzc{e}| <1 , \label{subluminal_image}
 \end{eqnarray}}
 with the possible exception of a countable set of points. Additionally, the protonic breaking points are mapped by a one-to-one correspondence onto the breaking points of the image trajectory according to 
 {\small
 \begin{eqnarray}
 \mathbf{x}_\mathpzc{p}(t^*_ \kappa)&=&  \mathbf{X}_\upmu (t^*_ \kappa + \mathpzc{\tau} ). \label{xalpha_sync}
 \end{eqnarray}}
 Notice that (\ref{xalpha_sync}) also relates every protonic breaking point $bp_\mathpzc{p}^{\kappa}$ to its \textit{mirror} electronic breaking point $bp_\mathpzc{e}^{\kappa} \equiv \textit{mirror}( bp_\mathpzc{p}^{ \kappa})$ whereby the breaking-point times are in the relation  {\small $(t_\mathpzc{p},t_\mathpzc{e}) \in \{ (t^*_\kappa, t^*_\kappa +\tau),  \, \kappa=1,\dots ,2N_{\flat}  \}$} with either $\mathpzc{\tau}=0, \,  \forall \kappa $ \,\, or \, \,$\mathpzc{\tau}=(T/2), \, \forall \kappa$. 
  \item[$(P3)$] At breaking points, the velocities jump discontinuously while satisfying 
  {\small
  \begin{eqnarray}
  \mathbf{v}_\mathpzc{p}^\ell(t^*_\kappa)&=&- \dot{\mathtt{G}}_\mathsmaller{B}(t^*_\kappa + \mathpzc{\tau}) + \upmu  \mathbf{v}_\mathpzc{e}^\ell(t^*_\kappa + \mathpzc{\tau}) \equiv \mathbf{V}_\upmu^\ell(t^*_\kappa + \mathpzc{\tau})  , \label{before1}\\  \mathbf{v}_\mathpzc{p}^\mathpzc{r}(t^*_\kappa)&=& - \dot{\mathtt{G}}_\mathsmaller{B}(t^*_\kappa + \mathpzc{\tau}) +\upmu  \mathbf{v}_\mathpzc{e}^\mathpzc{r}(t^*_\kappa + \mathpzc{\tau})  \equiv \mathbf{V}_\upmu^\mathpzc{r}(t^*_\kappa + \mathpzc{\tau})   \label{after1},
  \end{eqnarray}}
  where $\mathbf{V}_\upmu \equiv d\mathbf{X}_\upmu/dt$ and superscripts $\ell$ and $\mathpzc{r}$ indicate the velocity on the left-hand side and the velocity on the right-hand side of the breaking point, respectively.
 \end{enumerate}
 
The type-(ii) orbits further satisfying $P1$, $P2$ and $P3$ are henceforth called  \textit{musical orbits}.  The motivating example and prototype orbits to satisfy the musical properties are the spiky orbits illustrated in Figures 5 and 6 of Ref. \cite{cinderela}. 

\subsection{Synchronization function and distributional synchronization}
  \label{syncFu}
  For musical orbits, we further use the image trajectory (\ref{defimageXmu}) to construct a \textit{synchronization function} with an implicit state-dependent definition analogous to the definition of the delay functions (\ref{defi}), i.e.,
  {\small
\begin{eqnarray}
\phi_{\upmu}^{\pm}(t,\mathbf{x}) \equiv \vert \mathbf{x} -\mathbf{X}_{\upmu}(t \pm \phi_{\upmu}^{\pm})  \vert , 
\label{defimu}
\end{eqnarray}}
from where we define the corresponding \textit{synchronization times} by
{\small
\begin{eqnarray}
t_\upmu^\pm (t, \mathbf{x}) \equiv t \pm \phi_{\upmu}^{ \pm}(t,\mathbf{x}). \label{synctime}
\end{eqnarray} }
 Notice that since the image trajectory $\mathbf{X}_\upmu(t)$ is sub-luminal, the existence and the uniqueness of the synchronization functions (\ref{defimu}) are granted by theorem \ref{lema1}. A lower bound for the synchronization functions is obtained by replacing $\mathbf{x}_\mathpzc{j}(t)$ by $\mathbf{X}_\upmu(t)$ in theorem \ref{Fi_lower_bound}, yielding
 {\small
\begin{eqnarray}
\phi_{\upmu}^{\pm}(t,\mathbf{x}) \geq  |\mathbf{x}- \mathbf{X}_\upmu(t)| /2.
\label{Fi_alpha_lower_bound}
\end{eqnarray}}
Substituting $\mathbf{v}_\mathpzc{j}$ by $\mathbf{V}_{\upmu} \equiv d\mathbf{X}_{\upmu}/dt$ into formulas (\ref{gradient}) and (\ref{dfidt}), we obtain the gradient and time derivative of the synchronization functions (\ref{defimu}), i.e.,
{\small
\begin{eqnarray}
\vec{\nabla} \phi_\upmu^{\pm}(t,\mathbf{x})& =& \frac{\hat{\mathbf{n}}^\pm_\upmu}{(1 \pm \hat{\mathbf{n}}^\pm_\upmu \cdot \mathbf{V}^\pm_\upmu)} \label{gradientmu}, \\
\frac{ \partial \phi_\upmu^{\pm}}{\partial t}(t,\mathbf{x})& =& -\frac{ \hat{\mathbf{n}}^\pm_\upmu \cdot \mathbf{V}^\pm_\upmu}{(1 \pm \hat{\mathbf{n}}^\pm_\upmu \cdot \mathbf{V}^\pm_\upmu)} =\pm \left( |\vec{\nabla} \phi_\upmu^\pm| -1\right) \label{dfimudt},
\end{eqnarray}}
where the formula for $\hat{\mathbf{n}}_\upmu^\pm$ is obtained from (\ref{nj}) by replacing $\mathbf{x}_\mathpzc{j}$ with $\mathbf{X}_\upmu$, i.e., 
{\small
\begin{eqnarray}
\hat{\mathbf{n}}_\upmu^{\pm}(t,\mathbf{x}) \equiv \frac{ \mathbf{x} -\mathbf{X}_{\upmu}(t_\upmu^\pm) }{\phi_{\upmu}^{\pm}(t,\mathbf{x})}.
\label{nmu}
\end{eqnarray} }

 Next we show that the breaking image times  $t_\upmu^\pm(t,\mathbf{x})$ defined by (\ref{synctime}) are synchronized with the protonic breaking times $t_\mathpzc{p}^\pm(t, \mathbf{x})$. 
 \begin{lem} If ($P2$) holds, then for any $(t,\mathbf{x}) \in \mathbb{R} \times \mathbb{R}^3$, the deviating protonic time (\ref{light-cone_map})  is the time of a protonic breaking point, $t^* = t_\mathpzc{p}^\pm(t,\mathbf{x})$, \textit{if and only if} the deviating time (\ref{synctime}) is the time of a breaking point along the image trajectory,   $t_\upmu^\pm(t+\mathpzc{\tau},\mathbf{x})=t^{**}=t^{*}+\mathpzc{\tau}$. We also have $\hat{\mathbf{n}}_\mathpzc{p}^{\pm}(t^{*},\mathbf{x}) =\hat{\mathbf{n}}_{\upmu}^{ \pm}(t^{*}+ \mathpzc{\tau},\mathbf{x})$.
 \label{musicallema}
\end{lem}
\begin{proof} We start with the plus sign and fix $(t,\mathbf{x})$ in order for  $t^*=t_\mathpzc{p}^+ (t,\mathbf{x})$ to be the time of a protonic breaking point. Then, according to Eq. (\ref{xalpha_sync}), we have that $t^{**} = t^{*} +\mathpzc{\tau} $ is the time of a breaking point along the image trajectory. Using definition (\ref{light-cone_map}) with $\mathpzc{j}=\mathpzc{p}$, the one-to-one map (\ref{xalpha_sync}) of musical property ($P2$), and definition (\ref{synctime}) we have
 {\small
 \begin{eqnarray}
t^{**} =t+ \vert {{\mathbf{x} -\mathbf{x}}_{\mathpzc{p}}(t^*)} \vert +\mathpzc{\tau}= t +\mathpzc{\tau}+\vert {{\mathbf{x} - \mathbf{X}}_{\upmu}(t^{**})} \vert  =t
_\upmu^+ (t+\mathpzc{\tau}, \mathbf{x}) , \notag
\end{eqnarray}}
where in the last equality we have used definition (\ref{synctime}) with the plus sign. The result that  $\hat{\mathbf{n}}_\mathpzc{p}^+(t^{*},\mathbf{x}) =\hat{\mathbf{n}}_{\upmu}^+(t^{*}+\mathpzc{\tau},\mathbf{x})$ follows from musical property ($P2$) together with definitions (\ref{nj}) and (\ref{nmu}).  The proof for the minus-sign case is analogous.
\par
The converse part of the theorem is proved in the same way by exchanging the indices $\mathpzc{p}$ and $\upmu$ and exchanging $\tau$ by $-\tau$.
 \end{proof}
Our next result shows that the natural oscillatory functions
{\begin{eqnarray}
  \varDelta_\mathpzc{s}^\pm (t, \mathbf{x}) \equiv \phi_{\mathpzc{p}}^{\pm}(t,\mathbf{x}) - \phi_{\upmu}^{\pm}(t+\mathpzc{\tau},\mathbf{x}), \label{NATURAL_OSCILLATORY}
  \end{eqnarray}}
 and their first derivatives are continuous and bounded functions that vanish at breaking points.   
\begin{theorem} If ($P2$) and ($P3$) hold, the natural functions (\ref{NATURAL_OSCILLATORY}) and their first derivatives {\small $\vec{\nabla} \varDelta_\mathpzc{s}^\pm(t,\mathbf{x})$} and {\small $\partial_t \varDelta_\mathpzc{s}^\pm(t,\mathbf{x}) $}  are \textit{continuous}, bounded, and vanish at breaking points. \label{CONTI_BOUNDS}
 \end{theorem}
 \begin{proof} 
 \begin{enumerate}
 \item The natural functions (\ref{NATURAL_OSCILLATORY}) are continuous by (\ref{defi}) and (\ref{defimu}) because the trajectories are continuous. To prove that  $|\varDelta_\mathpzc{s}^\pm|$ is bounded we use (\ref{defi}), (\ref{defimu}) and the reverse triangular inequality, yielding
 {\small
 \begin{eqnarray}
|\varDelta_\mathpzc{s}^\pm(t,\mathbf{x})| &=&\bigg|  \vert { \mathbf{x} -\mathbf{x}_\mathpzc{p}^\pm} \vert  -\vert {\mathbf{x} - \mathbf{X}^\pm_\upmu  } \vert \bigg| \leq |\mathbf{x}_\mathpzc{p}^\pm -\mathbf{X}^\pm_\upmu | \leq \negthickspace \max _{t \in [0,T]} |\mathbf{x}_\mathpzc{p}(t)| +\negthickspace \max_{t \in [0,T]} |\mathbf{X}_\upmu(t) |,\notag \\ \label{invetri}
\end{eqnarray}}
and therefore the {\small $ \varDelta_\mathpzc{s}^\pm(t,\mathbf{x}) $} are bounded because the trajectories are bounded. The natural functions (\ref{NATURAL_OSCILLATORY})  vanish at breaking points because condition (\ref{xalpha_sync}) of musical property ($P2$) implies that $ \mathbf{X}^\pm_\upmu(t+\tau, \mathbf{x})=\mathbf{x}_\mathpzc{p}^\pm(t,\mathbf{x})$. 
\item  To prove continuity of the gradient of (\ref{NATURAL_OSCILLATORY})  we use (\ref{gradient}) and (\ref{gradientmu}) to yield
{\small
 \begin{eqnarray}
\vec{\nabla} \varDelta_\mathpzc{s}^\pm(t, \mathbf{x})=\left. \frac{\hat{\mathbf{n}}^\pm_\mathpzc{p}}{(1 \pm \hat{\mathbf{n}}_\mathpzc{p}^\pm \cdot \mathbf{v}^\pm_\mathpzc{p})}\right\vert_{(t,\mathbf{x})}  - \left. \frac{\hat{\mathbf{n}}^\pm_\upmu}{(1 \pm \hat{\mathbf{n}}^\pm_\upmu \cdot \mathbf{V}^\pm_\upmu)} \right\vert_{(t+\tau,\mathbf{x})} \label{gradvarfid}.
\end{eqnarray}}
We start from the plus sign case and drop the upper plus index to simplify the notation. According to Lemma \ref{musicallema}, for a fixed $\mathbf{x} \in \mathbb{R}^3$ each gradient is a function only of time \textit{and} the discontinuities happen only at the times $t^b(\mathbf{x})$ when $t^{**}=t_\upmu(t^b(\mathbf{x})+\mathpzc{\tau},\mathbf{x})=t_\mathpzc{p}(t^b(\mathbf{x}),\mathbf{x}) +\mathpzc{\tau}\equiv t^{*} +\tau$, at which times we have $\hat{\mathbf{n}}_\mathpzc{p}(t^{*},\mathbf{x})=\hat{\mathbf{n}}_\upmu(t^{*}+\mathpzc{\tau},\mathbf{x})$  (also by Lemma \ref{musicallema}). Conditions (\ref{before1}) and (\ref{after1}) of musical property $(P3)$ ensure that the gradient (\ref{gradvarfid}) \textit{vanishes} either from the left-hand side or from the right-hand side as $t$ crosses each breaking-point time $t^b(\mathbf{x})$, and therefore (\ref{gradvarfid}) is continuous at breaking points.  At all other times, both terms on the right-hand side of (\ref{gradvarfid}) are continuous and thus $\vec{\nabla} \varDelta_\mathpzc{s}^+(t, \mathbf{x})$ is continuous. The proof for the minus-sign case is analogous. The gradient (\ref{gradvarfid}) is bounded because the velocities are bounded along the sub-luminal periodic trajectories.
\item To prove that $\partial_t \varDelta_\mathpzc{s}^\pm(t, \mathbf{x})$ is continuous we use (\ref{dfidt}) and (\ref{dfimudt}) to yield
{\small
 \begin{eqnarray}
 \partial_t \varDelta_\mathpzc{s}^\pm (t, \mathbf{x})=\left. \frac{ \hat{\mathbf{n}}^\pm_\upmu  \cdot \mathbf{V}^\pm_\upmu}{(1 \pm \hat{\mathbf{n}}^\pm_\upmu \cdot \mathbf{V}^\pm_\upmu)}\right\vert_{(t+\tau,\mathbf{x})} -\left. \frac{\hat{\mathbf{n}}^\pm_\mathpzc{p} \cdot \mathbf{v}^\pm_\mathpzc{p}}{(1 \pm \hat{\mathbf{n}}^\pm_\mathpzc{p} \cdot \mathbf{v}^\pm_\mathpzc{p})}\right\vert_{(t,\mathbf{x})}  .\label{time_d}
\end{eqnarray}}
Again the proof is completed by noticing that for a fixed $\mathbf{x} \in \mathbb{R}^3 $ and at the times $t^b(\mathbf{x})$ such that  $t^*=t_\mathpzc{p} (t^b(\mathbf{x}),\mathbf{x})$ is a breaking-point along the protonic trajectory, we have by Lemma \ref{musicallema} that $\hat{\mathbf{n}}_\mathpzc{p}(t^b(\mathbf{x}),\mathbf{x}) =\hat{\mathbf{n}}_{\upmu}(t^b(\mathbf{x})+\mathpzc{\tau}, \mathbf{x})$ and conditions (\ref{before1}) and (\ref{after1}) of musical property ($P3$) ensure that the time-derivative (\ref{time_d}) \textit{vanishes} either from the left-hand side or from the right-hand side. The time derivative (\ref{time_d}) is bounded because the velocities are bounded.
\end{enumerate}
 \end{proof}
  In conformity with the Chemical Principle rationale, (\ref{NATURAL_OSCILLATORY}) describes the phase difference between the two electromagnetic waves acting on a third charge at $(t,\mathbf{x})$, as obtained by extending the light-cones of the third charge to reach the protonic and image trajectories. According to theorem \ref{lemmazero}, the singularities of the second derivatives (\ref{d2fidxx}) and (\ref{mixed_deriv}) are located where $\phi_\mathpzc{j}^\pm(t,\mathbf{x})=0$, which are \textit{on} trajectory $\mathpzc{j}$'s points by Lemma \ref{lemmazero}. Therefore, the singularities of $\nabla^2  \varDelta_\mathpzc{s}^\pm(t,\mathbf{x}) $ are located either \textit{on} the image trajectory (for $\mathpzc{j}=\upmu$) or \textit{on} the protonic trajectory (for $\mathpzc{j}=\mathpzc{p}$), and thus inside a ball $\mathcal{B}(|\mathbf{x}|< \mathpzc{r}_\mathsmaller{\pi}) \subset \mathbb{R}^3$ of (protonic) radius
 {\small
\begin{eqnarray}
\mathpzc{r}_\mathsmaller{\pi} \equiv \max_{t \in [0,T]} \{ |\mathbf{x}_\mathpzc{p}(t)|, |\mathbf{X_\upmu}(t)| \}. \label{radiusPP}
\end{eqnarray}}
The radius of the electronic trajectory does not appear in singular denominators and can be much larger than $\mathpzc{r}_{\mathsmaller{\pi}}$. The amplitude of the natural oscillation (\ref{NATURAL_OSCILLATORY}) is
{\small
\begin{eqnarray}
\rho_\mathpzc{os} \equiv \sup_{t \in \mathbb{R},\, \,  \mathbf{x} \in \mathbb{R}^3}  \{  | \varDelta_\mathpzc{s}^\pm (t, \mathbf{x})| \} \leq \max_{t \in [0,T]}  |\mathbf{x}_\mathpzc{p}(t)| +\max_{t \in [0,T]}  |\mathbf{X}_\upmu(t) | \leq 2\mathpzc{r}_{\mathsmaller{\pi}}, \label{AMPLIDELTA}
\end{eqnarray}}
where the first inequality is from theorem \ref{CONTI_BOUNDS}. 
\begin{theorem} For a musical orbit, the asymptotically vanishing combinations $\mathcal{R}_{\; \Delta}^\pm \equiv (\hat{\mathbf{r}} \cdot \vec{\nabla} \Delta^\pm_\mathpzc{s} \mp \partial_t \Delta_\mathpzc{s}^\pm)$ belong to  $\{ \mathbb{L}^\infty(\mathbb{R} \times \mathbb{R}^3)  \cap O(\frac{1}{r^2})\} $.\label{mystery_theorem} 
\end{theorem}
\begin{proof}
 We start from (\ref{gradient}) to express $ \hat{\mathbf{r}}\cdot \vec{\nabla} \Delta_\mathpzc{s}^\pm(t,\mathbf{x})$ as
{\small
\begin{eqnarray}
\negthickspace \negthickspace  \negthickspace \negthickspace \hat{\mathbf{r}}\cdot \vec{\nabla} \Delta_\mathpzc{s}^\pm&=&\left. |\nabla \phi_\mathpzc{p}^\pm| \right\vert_{(t,\mathbf{x})}-\left. |\nabla \phi_\upmu^\pm| \right\vert_{(t+\tau,\mathbf{x})}  \notag \\ &&+ \left. (\hat{\mathbf{n}}^\pm_\mathpzc{p}  \cdot \hat{\mathbf{r}} -1 ) |\nabla \phi_\mathpzc{p}^\pm|  \right\vert_{(t,\mathbf{x})} - \left. (\hat{\mathbf{n}}^+_\upmu  \cdot \hat{\mathbf{r}} -1 ) |\nabla \phi_\upmu^+|  \right\vert_{(t+\tau,\mathbf{x})}, \label{mystery}
\end{eqnarray}}
where we have added and subtracted $ ( |\nabla \phi_\mathpzc{p}^\pm|- |\nabla \phi_\upmu^\pm|)$, and re-arranged.  Using the last term of the right-hand side of (\ref{dfidt}) to evaluate the first line of the right-hand side of (\ref{mystery}), and using Eq. (\ref{expandnj}) to evaluate the second line of the right-hand side of (\ref{mystery}) yields
{\small
\begin{eqnarray}
\hat{\mathbf{r}}\cdot \vec{\nabla} \Delta^\pm_\mathpzc{s}(t,\mathbf{x})=\pm \partial_t \Delta^\pm_\mathpzc{s}(t,\mathbf{x}) +\mathcal{R}^\pm_{\; \Delta}(t,\mathbf{x}), \label{rgradphi}
\end{eqnarray}}
where we have defined the synchronized reminders $\mathcal{R}^\pm_{\; \Delta} (t,\mathbf{x}) $ to be
{\small
\begin{eqnarray}
\mathcal{R}^\pm_{\; \Delta} (t,\mathbf{x})  & \equiv & \left.   (\hat{\mathbf{n}}^\pm_\mathpzc{p}  \cdot \hat{\mathbf{r}} -1 ) |\nabla \phi_\mathpzc{p}^\pm|\right\vert_{(t,\mathbf{x})} - \left. (\hat{\mathbf{n}}^\pm_\upmu  \cdot \hat{\mathbf{r}} -1 ) |\nabla \phi_\upmu^\pm|  \right\vert_{(t+\tau,\mathbf{x})}=O(\frac{1}{r^2}). \label{errovarphiboth}   
\end{eqnarray}}
The first term on the right-hand side of (\ref{errovarphiboth}) shows that the $\mathcal{R}^\pm_{\; \Delta} (t,\mathbf{x})$ are bounded and the second identity was obtained using (\ref{expandnj}) to show that the $\mathcal{R}^\pm_{\; \Delta} (t,\mathbf{x})$ belong to  $\{ \mathbb{L}^\infty(\mathbb{R} \times \mathbb{R}^3)  \cap O(\frac{1}{r^2})\} $. 
\end{proof}
The importance of Eq. (\ref{rgradphi}) is that it transforms a space derivative into a time derivative plus an orbit-dependent  reminder belonging to $\{ \mathbb{L}^\infty(\mathbb{R} \times \mathbb{R}^3)  \cap O(\frac{1}{r^2})\} $. 

\begin{theorem}
 Let {\small $\mathcal{B}(|\mathbf{x}|< \mathpzc{r}_\mathsmaller{\pi}) \subset \mathbb{R}^3$} be a ball containing the protonic trajectory and the image trajectory (\ref{defimageXmu}). The natural functions (\ref{NATURAL_OSCILLATORY}) belong to $ \mathbb{W}^{\mathpzc{2},\mathpzc{2}}(\mathcal{B})$. 
\label{Sobolema}
\end{theorem}
\begin{proof}
\begin{enumerate}
\item The oscillatory functions (\ref{NATURAL_OSCILLATORY}) are locally integrable and thus define distributions $\mathfrak{D}^\pm$ on $\mathcal{B}$. Since the gradients of the $\varDelta_\mathpzc{s}^\pm(t,\mathbf{x})$ are continuous by theorem \ref{CONTI_BOUNDS}, the second distributional derivative of each distribution $\mathfrak{D}^\pm$ can be integrated piecewise by parts and because $(P3)$ holds each boundary term vanishes and the resulting regular distribution has a norm dominated by the sum of the Sobolev norms of $\phi_\mathpzc{p}(t,\mathbf{x})$ and $\phi_\upmu(t+\mathpzc{\tau},\mathbf{x})$. Otherwise the integration by parts generates a singular distribution.
\item  The Sobolev norm of $\varDelta_\mathpzc{s}^{\pm}$ is defined by 
{\small
\begin{eqnarray}
\negthickspace \negthickspace \negthickspace \negthickspace \negthickspace \negthickspace \negthickspace \negthickspace \negthickspace \negthickspace \negthickspace \negthickspace || \varDelta_\mathpzc{s}^{\pm} ||_{ \mathbb{W}^{\mathpzc{2},\mathpzc{2}}(\mathcal{B}) } \equiv \left(   \sum_{|\Bbbk| \leq 2} \int_{\mathcal{B}} (D^{\Bbbk} \varDelta_\mathpzc{s}^{\pm})^2 d^3 \mathbf{x}  \right)^{1/2}, \label{Bressan_norm_d}
\end{eqnarray}}
which involves a sum of integrals over $\mathcal{B}$ of squared partial derivatives for all multi-indices $\Bbbk$ satisfying $|\Bbbk| \leq 2$ \cite{Bressan}. The squared derivatives diverge either as $1/(\phi_\mathpzc{p} \phi_\mathpzc{p})$, $1/(\phi_\upmu \phi_\upmu)$ or as $1/(\phi_\mathpzc{p} \phi_\upmu)$ by Eqs. (\ref{d2fidxx}) and (\ref{mixed_deriv}), which are bounded according to  (\ref{COROBOUNDS}). The worst singularity is when $\mathbf{x}_\mathpzc{p}(t)=\mathbf{X}_\upmu(t+\mathpzc{\tau})$, at which times we use a coordinate system with origin at the common zero of $\phi_\mathpzc{p}(t,\mathbf{x})$ and $\phi_\upmu(t+\mathpzc{\tau},\mathbf{x})$ (i.e., $\mathbf{x}_{\mathpzc{O}}\equiv \mathbf{x}_\mathpzc{p}(t)= \mathbf{X}_\upmu(t+\mathpzc{\tau})$), whereby the integration volume $d^3 \mathbf{x}$ on $\mathcal{B}$ becomes proportional to the squared radius $r_{o}^2 \equiv  |\mathbf{x}- \mathbf{x}_\mathpzc{p}(t)|^2 $, and the integration over $\mathcal{B}$ is finite.  Using inequality (\ref{COROBOUNDS}) we find that all the integrals in (\ref{Bressan_norm_d}) are bounded because $\mathcal{B}$ is bounded, proving that $\varDelta_\mathpzc{s}^{\pm}(t,\mathbf{x}) \in \mathbb{W}^{\mathpzc{2},\mathpzc{2}}(\mathcal{B})$. For other values of $t$ we can divide the integration in two volumes separating the zeros of $\phi_\mathpzc{p}(t,\mathbf{x})$ and $\phi_\upmu(t+\mathpzc{\tau},\mathbf{x})$, and all integrations are bounded again by inequality (\ref{COROBOUNDS}).  
\end{enumerate}
\end{proof}

\section{Natural PDE and infinite-dimensional normed space} 
\label{PDE}

\subsection{Natural PDE defined from the musical orbit}
\label{natural_PDE}

The musical orbit defines {\small $\phi_\mathpzc{p}^\pm(t,\mathbf{x})$} and {\small $\phi_\mu^\pm(t,\mathbf{x})$} in $\mathbb{R}^4$ by theorem \ref{lema1} and the linear combinations $\varDelta_\mathpzc{s}^\pm(t,\mathbf{x})$ of theorem \ref{Sobolema} define natural PDEs by 
{\small
\begin{eqnarray}
  \negthickspace \negthickspace \negthickspace \negthickspace \negthickspace \negthickspace \negthickspace   \nabla^2  \varDelta_\mathpzc{s}^\pm(t,\mathbf{x}) & \equiv & \nabla^2 \phi_\mathpzc{p}^\pm(t,\mathbf{x}) -\nabla^2 \phi^\pm_\upmu(t+\tau,\mathbf{x}) \notag \\ &=&\left. |\vec{\nabla} \phi^\pm_\upmu|^3\hat{\mathbf{n}}^\pm_\upmu \cdot \mathbf{a}^\pm_\upmu \right\vert_{(t+\tau,\mathbf{x})}  -\left. |\vec{\nabla} \phi^\pm_\mathpzc{p}|^3\hat{\mathbf{n}}^\pm_\mathpzc{p} \cdot \mathbf{a}^\pm_\mathpzc{p} \right\vert_{(t,\mathbf{x})}  \notag \\ &&+ \left. \frac{(2|\vec{\nabla} \phi^\pm_\mathpzc{p}|+ |\vec{\nabla} \phi^\pm_\mathpzc{p}|^3 | \hat{\mathbf{n}}^\pm_\mathpzc{p} \times \mathbf{v}^\pm_\mathpzc{p} |^2) }{\phi^\pm_\mathpzc{p}}\right\vert_{(t,\mathbf{x})} \notag \\&& -\left. \frac{(2|\vec{\nabla} \phi^\pm_\upmu|+ |\vec{\nabla} \phi^\pm_\upmu|^3 | \hat{\mathbf{n}}^\pm_\upmu \times \mathbf{v}^\pm_\upmu |^2) }{\phi^\pm_\upmu} \right\vert_{(t+\tau,\mathbf{x})}.  \label{NATURAL2}
\end{eqnarray}}
On the last two lines of the right-hand side of (\ref{NATURAL2}), we have used (\ref{Laplacian_derivative}) to express the Laplacian derivatives of {\small $ \phi^\pm_\mathpzc{p}(t,\mathbf{x})$} and {\small $ \phi^\pm_\upmu(t+\tau,\mathbf{x})$}.
According to theorem \ref{Sobolema}, the natural oscillatory functions (\ref{NATURAL_OSCILLATORY}) belong to the Hilbert space $\mathbb{W}^{\mathpzc{2},\mathpzc{2}}(\mathcal{B})$, in which normed space first derivatives are \textit{continuous} by theorem \ref{CONTI_BOUNDS} and only the \textit{second} derivatives may be discontinuous. Equation (\ref{NATURAL2}) with the plus and the minus signs defines two PDEs involving second-derivative-only discontinuities. 
The following approximations are important to simplify the natural PDEs (\ref{NATURAL2}):

\begin{enumerate}
\item Using (\ref{expandphij}) in the region $|\mathbf{x}| \gg \mathpzc{r}_\mathsmaller{\pi}$ as defined by (\ref{radiusPP}), we can approximate the denominators $\phi_\mathpzc{j}^\pm(t,\mathbf{x})$ of the right-hand side of (\ref{NATURAL2}) by $\phi_\mathpzc{j}^\pm(t,\mathbf{x})  \simeq r $.
\item In the limit when the protonic radius (\ref{radiusPP}) goes to zero, the synchronization expected at breaking points by Lemma  \ref{musicallema} 
becomes global, i.e., 
{\small
\begin{eqnarray}
t_\mathpzc{p}^\pm(t,\mathbf{x}) & \rightarrow & (t \pm | \mathbf{x}|), \label{limit_tp} \\
t_\upmu^\pm(t+\tau,\mathbf{x}) & \rightarrow & (t +\tau \pm | \mathbf{x}|), \label{limit_tmu}
\end{eqnarray}}
which are \textit{explicitly} synchronized times. When the asymptotic limits (\ref{limit_tp}) and (\ref{limit_tmu}) hold, the oscillatory functions (\ref{NATURAL_OSCILLATORY}) are asymptotically periodic functions of the \textit{single} time $t_\upmu^\pm(t,\mathbf{x}) = \tau+ t_\mathpzc{p}^\pm =  t +\tau \pm |\mathbf{x}|$. Approximating the orbit locally by a harmonic oscillation, i.e., $\mathbf{x}_\mathpzc{p}(t_\mathpzc{p})  \propto  \vec{A}_\mathpzc{p}\cos(\mathpzc{i} \mathpzc{k}_{\; \mathpzc{q}} t_\mathpzc{p}) $ and $\mathbf{X}_\upmu (t_\upmu) \propto  \vec{A}_\upmu \cos(\mathpzc{i} \mathpzc{k}_{\; \mathpzc{q}} t_\upmu +\mathsmaller{\delta_\upmu} )$, we find using (\ref{expandphij}) that the four complex combinations $ \exp(\mp \mathpzc{i} \mathpzc{k}_{\; \mathpzc{q}} r) \varDelta_\mathpzc{s}^\pm(t,\mathbf{x})$ behave asymptotically as harmonic functions of time with coefficients depending on the direction $\hat{\mathbf{r}}$. The former suggests the inclusion of a compensating time oscillation by using the four products: $ \exp(- \mathpzc{i} (\mathpzc{k}_{\; \mathpzc{q}} r+\varpi_\mathpzc{q}t) ) \varDelta_\mathpzc{s}^+(t,\mathbf{x})$, $ \exp(- \mathpzc{i} (\mathpzc{k}_{\; \mathpzc{q}} r-\varpi_\mathpzc{q}t) ) \varDelta_\mathpzc{s}^+(t,\mathbf{x})$, $ \exp( \mathpzc{i} (\mathpzc{k}_{\; \mathpzc{q}} r-\varpi_\mathpzc{q}t) ) \varDelta_\mathpzc{s}^-(t,\mathbf{x})$ and $ \exp( \mathpzc{i} (\mathpzc{k}_{\; \mathpzc{q}} r+\varpi_\mathpzc{q}t) ) \varDelta_\mathpzc{s}^-(t,\mathbf{x})$. The meaning of the compensation within the Chemical Principle rationale is that of waves coming to and from the third particle.
\end{enumerate}
 As discussed in the above item (1.),  we can use (\ref{expandphij}) to approximate the rotating singularities by an effective singularity proportional to $\frac{1}{r}$ plus an $O(\frac{1}{r^2})$ reminder $\mathpzc{R}^\pm_{\,\,\mathpzc{p} \upmu}(t,\mathbf{x})$ and thus express Eq. (\ref{NATURAL2}) as 
{\small
\begin{eqnarray}
\nabla^2 \varDelta_\mathpzc{s}^\pm(t,\mathbf{x})
& =& -\mathpzc{A}_{ \, \, \mathpzc{p} \upmu }^\pm(t,\mathbf{x})+ \frac{(\mathpzc{Q}_{\, \, \mathpzc{p}}^\pm(t,\mathbf{x}) -\mathpzc{Q}_{\, \, \upmu}^\pm(t,\mathbf{x}) ) }{r} +\;  \mathpzc{R}^\pm_{\,\,\mathpzc{p} \upmu}(t,\mathbf{x}).\notag \\ \label{regular2R}
\end{eqnarray}}
The right-hand side of Eq. (\ref{regular2R}) is expressed in terms of $\mathpzc{Q}_{\, \, \mathpzc{p}}^\pm(t,\mathbf{x}) \in \mathbb{L}^{ \infty} (\mathbb{R}^3)$, $\mathpzc{Q}_{\, \, \upmu}^\pm(t,\mathbf{x}) \in \mathbb{L}^{ \infty} (\mathbb{R}^3) $ and $\mathpzc{A}_{ \, \,\mathpzc{p} \upmu}^\pm(t,\mathbf{x}) \in \mathbb{L}^{ \infty} (\mathbb{R}^3)$, which are defined respectively by
{\small
\begin{eqnarray}
\mathpzc{Q}_{\, \, \mathpzc{p}}^\pm(t,\mathbf{x}) &\equiv & \left. ( 2 |\vec{\nabla} \phi^\pm_\mathpzc{p}| +| \vec{\nabla} \phi^\pm_{\mathpzc{p}} |^3 |\hat{\mathbf{n}}^\pm_{\mathpzc{p}} \times \mathbf{v}^\pm_{\mathpzc{p}} |^2 ) \right\vert_{(t,\mathbf{x})},  \label{Qone_p} \\
\mathpzc{Q}_{\, \, \upmu}^\pm(t,\mathbf{x}) &\equiv & \left. ( 2 |\vec{\nabla} \phi^\pm_\upmu| +| \vec{\nabla} \phi^\pm_{\upmu} |^3 |\hat{\mathbf{n}}^\pm_{\upmu} \times \mathbf{V}^\pm_{\upmu} |^2 ) \right\vert_{(t+\mathpzc{\tau},\mathbf{x})},  \label{Qone_mu} \\
\mathpzc{A}_{ {\, \, \mathpzc{p} \upmu} }^\pm(t,\mathbf{x}) & \equiv& \left. | \vec{\nabla} \phi^\pm_{\mathpzc{p}} |^3\hat{\mathbf{n}}^\pm_\mathpzc{p} \cdot \mathbf{a}^\pm_\mathpzc{p} \right\vert_{(t,\mathbf{x})} - \left. | \vec{\nabla} \phi^\pm_{\upmu} |^3 \hat{\mathbf{n}}^\pm_{\upmu} \cdot \mbox{\Large $a$}_\upmu^\pm \right\vert_{(t+\mathpzc{\tau},\mathbf{x})},\label{NEW_KL} 
\end{eqnarray}}
where $\mbox{\Large $a$}_\upmu^\pm (t,\mathbf{x}) $ is the acceleration of the image trajectory (\ref{defimageXmu}). In Eq. (\ref{regular2R}) the reminders $\mathpzc{R}_{\,\, \mathpzc{p} \upmu}^\pm (t,\mathbf{x}) \in \{ \mathbb{L}^2(\mathbb{R}^3) \cap O(\frac{1}{r^2}) \}$ are defined by
{\small
\begin{eqnarray}
\negthickspace \mathpzc{R}_{\,\, \mathpzc{p} \upmu}^\pm  \equiv  \bigg( \, \frac{1}{\phi^\pm_\mathpzc{p}(t,\mathbf{x})}- \frac{1}{r}\, \bigg) \mathpzc{Q}_{\, \, \mathpzc{p}}^\pm(t,\mathbf{x})-\bigg(\, \frac{1}{\phi^\pm_\upmu(t+\tau,\mathbf{x})}- \frac{1}{r} \bigg)\mathpzc{Q}_{\, \, \upmu}^\pm(t,\mathbf{x}). \label{resto_K}
\end{eqnarray}}
\par
 As suggested by the limit discussed in the above item (2.), we shall include a wave to compensate the asymptotic oscillatory behaviour of the $\Delta_\mathpzc{s}^\pm$. From the \textit{four} possible compensated products, here we use only two products in order to construct our two-component functions, i.e., we define $\varphi : \, (t, \mathbf{x}) \in \mathbb{R}\times \mathbb{R}^3 \rightarrow \mathbb{C}$ and $\varphi^\dagger : \, (t, \mathbf{x}) \in \mathbb{R}\times \mathbb{R}^3 \rightarrow \mathbb{C}$ by
 {\small
 \begin{eqnarray}
 \varphi(t,\mathbf{x}) &\equiv& 
    \alpha \exp(-\mathpzc{i} \mathpzc{k}_{\;\mathsmaller{q}} r-\mathpzc{i} \varpi_\mathpzc{q} t) \varDelta_\mathpzc{s}^+(t,\mathbf{x})  + \beta \exp(\mathpzc{i} \mathpzc{k}_{\; \mathpzc{q}} r -\mathpzc{i} \varpi_\mathpzc{q} t) \varDelta_\mathpzc{s}^-(t,\mathbf{x}), \label{rotating} \\
\varphi^\dagger (t,\mathbf{x}) &\equiv& 
    \alpha \exp(-\mathpzc{i} \mathpzc{k}_{\; \mathpzc{q}} r-\mathpzc{i} \varpi_\mathpzc{q} t) \varDelta_\mathpzc{s}^+(t,\mathbf{x})  - \beta \exp(\mathpzc{i} \mathpzc{k}_{\; \mathpzc{q}} r-\mathpzc{i} \varpi_\mathpzc{q} t) \varDelta_\mathpzc{s}^-(t,\mathbf{x}). \label{dagger} 
\end{eqnarray} }
In Eqs. (\ref{rotating}) and (\ref{dagger}), the real $\mathpzc{k}_{\; \mathpzc{q}}$ is the wave number and the real $\varpi_\mathpzc{q}$ is the wave frequency. The complex numbers $\alpha$ and $\beta$ in Eq. (\ref{dagger}) are henceforth called spinorial components and the upper dagger in $\varphi^\dagger(t,\mathbf{x})$ indicates the function obtained from $\varphi(t,\mathbf{x})$ by replacing $\beta \rightarrow -\beta$, not to be confused with the complex conjugate. The gradient of $\varphi(t,\mathbf{x})$ is
{\small
\begin{eqnarray}
\negthickspace \negthickspace  \negthickspace \negthickspace \vec{\nabla} \varphi(t,\mathbf{x})\negthickspace \negthickspace &=&\negthickspace \negthickspace \negthickspace -\mathpzc{i} \mathpzc{k}_{\; \mathpzc{q}}\varphi^\dagger(t,\mathbf{x})
\hat{\mathbf{r}} \notag \\ && +  \alpha \exp{(-\mathpzc{i} \mathpzc{k}_{\; \mathpzc{q}} r-\mathpzc{i} \varpi_\mathpzc{q} t)} \vec{\nabla} \varDelta_\mathpzc{s}^+(t,\mathbf{x}) +\beta  \exp{(\mathpzc{i} \mathpzc{k}_{\; \mathpzc{q}} r -\mathpzc{i} \varpi_\mathpzc{q} t)}\vec{\nabla} \varDelta_\mathpzc{s}^-(t,\mathbf{x}),   \label{gradquasi}
\end{eqnarray}}
where $\hat{\mathbf{r}} \equiv \frac{\mathbf{x}}{|\mathbf{x}|}$ and we have used (\ref{dagger}). The Laplacian derivative of $\varphi(t,\mathbf{x})$ can be evaluated using (\ref{regular2R}), (\ref{rotating}) and theorem \ref{mystery_theorem}. Disregarding the reminders $\mathpzc{R}_{\,\, \mathpzc{p} \upmu}^\pm (t,\mathbf{x}) \in \{ \mathbb{L}^2(\mathbb{R}^3) \cap O(\frac{1}{r^2}) \}$ and $\mathcal{R}^\pm_{\; \Delta} (t,\mathbf{x}) \in \{ \mathbb{L}^\infty(\mathbb{R} \times \mathbb{R}^3)  \cap O(\frac{1}{r^2})\} $ we have
{\small
\begin{eqnarray}
 \nabla^2 \varphi&\equiv&-2\delta \mathpzc{a}_{ \mathpzc{p} \upmu }   +(\mathpzc{k}^2_{\; \mathpzc{q}}+ 2\mathpzc{k}_{\; \mathsmaller{D}}\mathpzc{k}_{\; \mathpzc{q}})\varphi -2\mathpzc{i} \mathpzc{k}_{\; \mathpzc{q}} \partial_t \varphi  + \frac{2}{r} \bigg( \delta \mathpzc{v}_{ \mathpzc{p} \upmu}+( \partial_t \varphi^\dagger   -\mathpzc{i}\mathpzc{k}_{\; \mathsmaller{D}}  \varphi^\dagger)   \bigg),  \label{SOBRE_NATURAL}
 \end{eqnarray}
}
where the real number {\small $\mathpzc{k}_{\; \mathsmaller{D}}$} is defined by 
{\small
\begin{eqnarray}
\mathpzc{k}_{\; \mathsmaller{D}} \equiv \mathpzc{k}_{\; \mathpzc{q}} -\varpi_{\mathpzc{q}}, \label{defKDBROG}
\end{eqnarray}
}
and we have used (\ref{Qone_p}), (\ref{Qone_mu}) and (\ref{NEW_KL}) to define some auxiliary quantities from the coefficients $\mathpzc{A}_{ \, \, \mathpzc{p} \upmu }^\pm$ and $(\mathpzc{Q}_{\, \, \mathpzc{p}}^\pm -\mathpzc{Q}_{\, \, \upmu}^\pm) $ of (\ref{regular2R}) as
 {\small
\begin{eqnarray}
 \negthickspace \negthickspace  \negthickspace \negthickspace \negthickspace \negthickspace  \delta \mathpzc{v}_{ \mathpzc{p} \upmu}(t,\mathbf{x}) &\equiv & \frac{\alpha}{2} \exp(-\mathpzc{i} \mathpzc{k}_{\; \mathpzc{q}} r-\mathpzc{i} \varpi_\mathpzc{q} t)    \left(   \left.  | \vec{\nabla} \phi_\mathpzc{p}^+ |^3 |\hat{\mathbf{n}}_\mathpzc{p}^+ \times \mathbf{v}_\mathpzc{p}^+ |^2 \right\vert_{(t,\mathbf{x})} -\left.  | \vec{\nabla} \phi_\upmu^+ |^3 |\hat{\mathbf{n}}_\upmu^+ \times \mathbf{V}_\upmu^+ |^2\right\vert_{(t+\mathpzc{\tau},\mathbf{x})} \right)\notag \\ && \negthickspace \negthickspace \negthickspace +\frac{ \beta}{2} \exp(\mathpzc{i} \mathpzc{k}_{\; \mathpzc{q}} r-\varpi_\mathpzc{q} t)\left(  \left. | \vec{\nabla} \phi_\mathpzc{p}^- |^3 |\hat{\mathbf{n}}_\mathpzc{p}^- \times \mathbf{v}_\mathpzc{p}^- |^2 \right\vert_{(t,\mathbf{x})} -\left. | \vec{\nabla} \phi_\upmu^- |^3 |\hat{\mathbf{n}}_\upmu^- \times \mathbf{V}_\upmu^- |^2 \right\vert_{(t+\mathpzc{\tau},\mathbf{x})} \right), \notag \\
  \label{epsonV} \\
 \negthickspace \negthickspace  \negthickspace \negthickspace \negthickspace \negthickspace \delta \mathpzc{a}_{ \mathpzc{p} \upmu } (t,\mathbf{x}) &\equiv & \frac{\alpha}{2} \exp{(-\mathpzc{i} \mathpzc{k}_{\; \mathpzc{q}} r-\mathpzc{i}\varpi_\mathpzc{q} t)} \left(   \left. | \vec{\nabla} \phi^+_{\mathpzc{p}} |^3\hat{\mathbf{n}}^+_\mathpzc{p} \cdot \mathbf{a}^+_\mathpzc{p} \right\vert_{(t,\mathbf{x})}  -\left. | \vec{\nabla} \phi^+_{\upmu} |^3 \hat{\mathbf{n}}^+_{\upmu} \cdot \mbox{\Large $a$}_\upmu^+ \right\vert_{(t+\mathpzc{\tau},\mathbf{x})}  \right) \notag \\
 & & \negthickspace \negthickspace \negthickspace + \frac{\beta}{2} \exp{(\mathpzc{i} \mathpzc{k}_{\; \mathpzc{q}} r-\mathpzc{i}\varpi_\mathpzc{q} t)} \left(    \left. | \vec{\nabla} \phi^-_{\mathpzc{p}} |^3\hat{\mathbf{n}}^-_\mathpzc{p} \cdot \mathbf{a}^-_\mathpzc{p} \right\vert_{(t,\mathbf{x})}  -\left. | \vec{\nabla} \phi^-_{\upmu} |^3 \hat{\mathbf{n}}^-_{\upmu} \cdot \mbox{\Large $a$}_\upmu^- \right\vert_{(t+\mathpzc{\tau},\mathbf{x})} \right),
 \label{epsonA} 
\end{eqnarray} }
with $\mbox{\Large $a$}_\upmu^\pm (t,\mathbf{x}) $ being the acceleration of the image trajectory (\ref{defimageXmu}). 

\subsection{Qualitative classification of the reminders}
\label{estima_e_classifica}
 The reminders $ \mathpzc{R}_{\,\, \mathpzc{p} \upmu}^\pm  $ and $\mathcal{R}^\pm_{\; \Delta} $ vanish at breaking points just like the quantities of theorem \ref{CONTI_BOUNDS}, as follows. The vanishing of $ \mathpzc{R}_{\,\, \mathpzc{p} \upmu}^\pm  $ at breaking points follows from (\ref{resto_K}) using properties ($P2$) and ($P3$) and Eqs. (\ref{Qone_p}) and (\ref{Qone_mu}). The vanishing of $\mathcal{R}^\pm_{\; \Delta} $ at breaking points follows from (\ref{errovarphiboth}) together with properties ($P2$) and ($P3$). The synchronization of the zeros of $ \mathpzc{R}_{\,\, \mathpzc{p} \upmu}^\pm  $ and $\mathcal{R}^\pm_{\; \Delta} $ with the zeros of the natural quantities of theorem \ref{CONTI_BOUNDS} is \textit{asking} for an approximation by linear functions of $\Delta_\mathpzc{s}^\pm$, $\partial_t \Delta_\mathpzc{s}^\pm$ and $\vec{\nabla}\Delta_\mathpzc{s}^\pm$, henceforth a \textit{linear bridging} approximation. An emblematic example in the class of linear bridging approximations for the asymptotic tail of $\mathcal{R}^\pm_{\; \Delta}(t,\mathbf{x}) $ is  the approximation by linear functions of $|\hat{\mathbf{r}}\times \vec{\nabla} \Delta^\pm_\mathpzc{s}| $ obtained using (\ref{expandrtimesnj}), yielding  
{\small
\begin{eqnarray}
 \mathcal{R}^\pm_{\; \Delta} (t,\mathbf{x}) \equiv \left\{ \begin{array}{ll}
      \mathpzc{R}^\pm_{\; \mathsmaller{\Delta , \, opt.}}(t,\mathbf{x}) \in \{ \mathbb{L}^\infty (\mathbb{R} \times \mathbb{R}^3) \cap \mathbb{L}^2(\mathbb{R}^3) \} \;  \ \mbox{if} \; \; r \leq \mathpzc{r}_\mathsmaller{spin},\\
      \frac{\mathpzc{S}_1^\pm |\hat{\mathbf{r}}\times \vec{\nabla} \Delta^\pm_\mathpzc{s}| }{r^2}\exp{(\mathpzc{i} \varpi_\mathpzc{o} t-\mathpzc{q} r)} + O(\frac{\Delta^\pm_\mathpzc{s}}{r^3}) \; \;  \mbox{if} \; \; r > \mathpzc{r}_\mathsmaller{spin},
    \end{array}  \right.
    \label{spin-orbit}
\end{eqnarray}}
where the real $\mathpzc{r}_\mathsmaller{spin}>0$ and complex $(\mathpzc{S}_1^+, \mathpzc{S}_1^-)$ are optimal numbers and  $\mathpzc{R}^\pm_{\; \mathsmaller{\Delta , \, opt.}}(t,\mathbf{x})$ are functions with a minimal norm. On the second line of (\ref{spin-orbit}), the symbol $O(\frac{\Delta^\pm_\mathpzc{s}}{r^3})$ indicates that the tail vanishes when $\Delta_\mathpzc{s}^\pm(t,\mathbf{x})$ vanishes. If expressed in terms of $\Psi(t,\mathbf{x})$ and $\Psi^\dagger(t,\mathbf{x})$, the asymptotic tail of (\ref{spin-orbit}) has the exact functional form of the \textit{spin-orbit term} of quantum mechanics\cite{Salpeter}. We henceforth call  $\mathpzc{R}_{\; \mathsmaller{ID}}$, $ \mathpzc{R}_{\,\, \mathpzc{p} \upmu}^\pm  $ and $\mathcal{R}^\pm_{\; \Delta} $ spin-orbit reminders \textit{because} of the suitability of approximation (\ref{spin-orbit}).

 \subsection{From a PDE in $\mathbb{W}^{\mathpzc{2},\mathpzc{2}}(\mathcal{B})$ to a PDE in $\mathbb{W}^{\mathpzc{2},\mathpzc{2}}(\mathbb{R}^3)$}
\label{linear_PDE}

The natural PDE (\ref{regular2R}) can be extended to $\mathbb{H}^2 \equiv \mathbb{W}^{\mathpzc{2},\mathpzc{2}}(\mathbb{R}^3)$ by using polynomial combinations of $\varphi  \in \mathbb{W}^{\mathpzc{2},\mathpzc{2}}(\mathcal{B})$ and $\varphi^\dagger  \in \mathbb{W}^{\mathpzc{2},\mathpzc{2}}(\mathcal{B})$ defined in (\ref{rotating}) and (\ref{dagger}) with coefficients that are decreasing exponentials of $r \equiv |\mathbf{x}|$. From (\ref{rotating}) and (\ref{dagger}) we further define the square-normalizable complex function $\Psi : \, (t, \mathbf{x}) \in \mathbb{R}\times \mathbb{R}^3 \rightarrow \mathbb{C}$ and the square-normalizable complex function $\Psi^\dagger : \, (t, \mathbf{x}) \in \mathbb{R}\times \mathbb{R}^3 \rightarrow \mathbb{C}$ by
{\small
\begin{eqnarray}
\Psi (t, \mathbf{x}) &\equiv& \mathcal{P} (\varphi)  \exp( \mathpzc{i} \varpi_\mathpzc{o} t-\mathpzc{q} r) \label{defPSI},\\
\Psi^\dagger (t, \mathbf{x}) &\equiv& \mathcal{P} (\varphi^\dagger)   \exp( \mathpzc{i} \varpi_\mathpzc{o} t-\mathpzc{q} r), \label{defPSI_dagger}
\end{eqnarray}}
where $\mathpzc{q} > 0$ is real, \;$\varpi_\mathpzc{o}$ is real and $\mathcal{P} (\varphi)$ is a quasi-polynomial of the variable $\varphi$. In Eq. (\ref{defPSI_dagger}), $\Psi^\dagger(t,\mathbf{x})$ indicates the function obtained from $\Psi(t,\mathbf{x})$ by replacing $\beta \rightarrow -\beta$, again, not to be confused with the complex conjugate. The functions $\Psi (t, \mathbf{x})$ and $\Psi^\dagger (t, \mathbf{x})$ defined by Eqs. (\ref{defPSI}) and (\ref{defPSI_dagger}) inherit a Laplacian derivative and a continuous gradient defined almost everywhere because $\varphi (t, \mathbf{x})$ and $\varphi^\dagger (t, \mathbf{x})$ possess these properties by theorem \ref{CONTI_BOUNDS}.  The second-derivatives of (\ref{defPSI}) and (\ref{defPSI_dagger}) introduces again the same $\frac{1}{r}$ singularity because $\nabla^2 r=\frac{2}{r}$.
\begin{theorem}
 For a musical orbit, the functions (\ref{defPSI}) and (\ref{defPSI_dagger}) belong to $ \mathbb{W}^{\mathpzc{2},\mathpzc{2}}(\mathbb{R}^3)$ for any $\mathpzc{q}>0$.
\label{Sobolema3}
\end{theorem}
\begin{proof}
\begin{enumerate}
\item The derivatives of $\Psi(t,\mathbf{x})$ are easily evaluated from (\ref{defPSI}). The first derivative respect to $x$ is
{\small
\begin{eqnarray}
\frac{\partial \Psi}{\partial x} = \left( - \mathpzc{q} {\hat{\mathbf{r}}}_x + \mathcal{P}^\prime(\varphi)  \frac{\partial \varphi}{\partial x} \right) \exp( \mathpzc{i} \varpi_\mathpzc{o} t-\mathpzc{q} r),
\label{gradPhi}
\end{eqnarray}}
where ${\hat{\mathbf{r}}}_x \equiv x/r $ and analogous expressions hold for the partial derivatives respect to $y$ and the $z$. The second derivatives respect to $xx$ and $xy$ are
{\small
\begin{eqnarray}
\frac{\partial^2 \Psi}{\partial x^2} &=& \left( (\mathpzc{q}^2 {\hat{\mathbf{r}}}_x^2 -\mathpzc{q}\frac{\partial {\hat{\mathbf{r}}}_x }{\partial x}) \mathcal{P}(\varphi) + (\frac{\partial \varphi}{\partial x})^2 \mathcal{P}^{\prime \prime} (\varphi)  \right)   \exp(\mathpzc{i} \varpi_\mathpzc{o} t -\mathpzc{q}r) \notag \\ & &+ \mathcal{P}^\prime(\varphi)\left(\frac{\partial^2 \varphi}{\partial x^2} -2\mathpzc{q} {\hat{\mathbf{r}}}_x  \frac{\partial \varphi}{\partial x}\right)   \exp{(\mathpzc{i} \varpi_\mathpzc{o} t-\mathpzc{q}r)},   \label{Phixx} \\
\frac{\partial^2 \Psi}{\partial x \partial y} &=&  \left( (\mathpzc{q}^2 {\hat{\mathbf{r}}}_x {\hat{\mathbf{r}}}_y -\mathpzc{q}\frac{\partial {\hat{\mathbf{r}}}_x }{\partial y}) \mathcal{P}(\varphi) + (\frac{\partial \varphi}{\partial x})(\frac{\partial \varphi}{\partial y})  \mathcal{P}^{\prime \prime} (\varphi)  \right)   \exp(\mathpzc{i} \varpi_\mathpzc{o} t-\mathpzc{q}r) \notag \\ & &+ \mathcal{P}^\prime(\varphi)\left(\frac{\partial^2 \varphi}{\partial x^2} -\mathpzc{q} ( {\hat{\mathbf{r}}}_x  \frac{\partial \varphi}{\partial y}  + {\hat{\mathbf{r}}}_y  \frac{\partial \varphi}{\partial x}) \right)   \exp(\mathpzc{i} \varpi_\mathpzc{o} t-\mathpzc{q}r)  , 
\label{Phixy}\end{eqnarray}}
where $\frac{ \partial {\hat{\mathbf{r}}}_x}{\partial x}  \equiv \frac{(1-{\hat{\mathbf{r}}}_x^2)}{r}$ and $\frac{ \partial {\hat{\mathbf{r}}}_x}{\partial y}  \equiv -\frac{ {\hat{\mathbf{r}}}_x  {\hat{\mathbf{r}}}_y}{r}$. The other partial derivatives are obtained by permuting $x$, $y$ and $z$ in the above formulas.
 \item We generalize the Sobolev norm for complex functions by
 {\small
\begin{eqnarray}
|| \Psi ||_{ \mathbb{W}^{\mathpzc{2},\mathpzc{2}}(\mathbb{R}^3) } \equiv \left(   \sum_{|\Bbbk| \leq 2} \int_{\mathbb{R}^3} (D^{\Bbbk} \Psi)(D^{\Bbbk} \Psi)^{*} d^3 \mathbf{x}  \right)^{1/2}, \label{Bressan_norm2}
\end{eqnarray}}
where the upper star indicates complex conjugation and $\Bbbk$ is the multi-index of the partial derivative \cite{Bressan}. As proved in theorem \ref{CONTI_BOUNDS}, the $ \varDelta_\mathpzc{s}^\pm(t, \mathbf{x})$ are bounded and it can be seen from (\ref{gradquasi}) that the modulus $| \vec{\nabla} \varphi|$ is bounded and the second derivatives  $\frac{\partial^2 \varphi}{\partial x \partial y}$ and $\frac{\partial^2 \varphi}{\partial x^2}$ diverge at the most as {\small $\frac{1}{r}$} or {\small $\frac{1}{\phi_\mathpzc{j}}$} according to Eqs. (\ref{gradient}), (\ref{d2fidxx}) and (\ref{mixed_deriv}), thus belonging to $\mathbb{L}^2_{loc}(\mathbb{R}^3)$ by  inequality (\ref{COROBOUNDS}). The former is sufficient for the decreasing exponential on (\ref{defPSI}), (\ref{gradPhi}), (\ref{Phixx})  and (\ref{Phixy}) to dominate the quasi-polynomials and ensure that $|\Psi(t,\mathbf{x})|^2$ and its squared derivatives up to order two belong to $\mathbb{L}^2(\mathbb{R}^3)$, as necessary for the norm (\ref{Bressan_norm2}) to be finite and $\Psi(t,\mathbf{x}) \in \mathbb{H}^2 \equiv \mathbb{W}^{\mathpzc{2},\mathpzc{2}}(\mathbb{R}^3)$. The details about integrating the divergencies are the same outlined in theorems \ref{Sobolema} and \ref{Sobolema3}, and the proof that $\Psi^\dagger (t, \mathbf{x}) \in  \mathbb{W}^{\mathpzc{2},\mathpzc{2}}(\mathbb{R}^3)$ is the same. 
\end{enumerate}
\end{proof}

\subsection{Operator identity for the musical orbit}
\label{identity}
From a musical orbit we define the set $ \mathcal{A}_{(ii)} \subset \mathbb{W}^{\mathpzc{2},\mathpzc{2}}(\mathbb{R}^3)$ of all functions $\Psi(t,\mathbf{x})$ of the form (\ref{defPSI}).  Since the $\Psi(t,\mathbf{x})$ belongs to $ \mathbb{W}^{\mathpzc{2},\mathpzc{2}}(\mathbb{R}^3)$ by theorem \ref{Sobolema3}, we define the Schroedinger linear operator $\mathcal{S} :  \Psi(t,\mathbf{x}) \in \mathcal{A}_{(ii)} \subset \mathbb{W}^{\mathpzc{2},\mathpzc{2}}(\mathbb{R}^3)  \rightarrow \mathbb{L}^2(\mathbb{R}^3)$ by
{\small
 \begin{eqnarray}
 \mathcal{S}(\Psi(t,\mathbf{x}) ) \equiv  \frac{1}{2}\mathpzc{r}_{\mathsmaller{B}} \nabla^2 \Psi + \frac{1}{r} \Psi  + \mathpzc{i}\mathpzc{h}_{\mathsmaller{B}} \partial_t \Psi ,  \label{APDE}
 \end{eqnarray}}
 where $\mathpzc{r}_{\mathsmaller{B}}>0$ and $\mathpzc{h}_\mathsmaller{B} $ are reals, $r \equiv | \mathbf{x}  | $ and $\mathpzc{i} \equiv \sqrt{-1}$ is the complex unit. We henceforth assume that  $\mathpzc{r}_\mathsmaller{B}$ satisfies the inequality
{\small
\begin{eqnarray}
\negthickspace \negthickspace \negthickspace \negthickspace \negthickspace \negthickspace \negthickspace \negthickspace \negthickspace \negthickspace \mathpzc{k}_{\; \lambda}   & \equiv &  \frac{1}{\mathpzc{r}_\mathsmaller{B} } -\mathpzc{q} > 0.  
\label{definelamB} 
\end{eqnarray}}
Notice on the right-hand side of (\ref{APDE}) that the following quantities belong to  $\mathbb{L}^2(\mathbb{R}^3)$: 
 \begin{itemize}
 \item  $\partial_t \Psi$ is a continuous function because $\partial_t \varDelta_\mathpzc{s}^\pm $ is continuous by theorem \ref{CONTI_BOUNDS} and moreover $\partial_t \Psi \in \mathbb{L}^2(\mathbb{R}^3)$  because it is a quasi-polynomial of the bounded oscillatory function (\ref{rotating}) times a decreasing exponential.
 \item The second term on the right-hand side of (\ref{APDE}), (namely $\frac{\Psi}{r}$), belongs to $\mathbb{L}^2(\mathbb{R}^3)$ because the integration element $d^3 \mathbf{x}=4\pi r^2 d \hat{\omega}_{\mathsmaller{2}}$ of $\mathbb{R}^3$ cancels the $\frac{1}{r^2}$ factor and the remaining $|\Psi(t,\mathbf{x})|^2$ is integrable because the decreasing exponential dominates any quasi-polynomial.
 \item $\nabla^2 \Psi $ belongs to  $\mathbb{L}^2(\mathbb{R}^3)$ by theorem \ref{Sobolema3} and it is the only term that can possibly be discontinuous on the right-hand side of (\ref{APDE}). In the case $\mathcal{S}=0$, then $\nabla^2 \Psi $ must be continuous as well, because the other two terms of the right-hand side of (\ref{APDE}) are continuous.
 \end{itemize}
 Since the singularities of $\mathcal{S}(\Psi)$ are located \textit{on} the protonic and image trajectories by theorem \ref{lemmazero}, the $\mathbb{L}^2(\mathbb{R}^3)$ norm of $\mathcal{S}(\Psi)$ measures the effective influence of the rotating singularities and, by varying the real parameters   $\mathpzc{r}_\mathsmaller{B}>0$ and $\mathpzc{h}_\mathsmaller{B}$ of (\ref{APDE}), one can minimize $S(\Psi)$ in $\mathbb{L}^2(\mathbb{R}^3)$ to obtain an \textit{effective} linear Schroedinger equation. In order to evaluate the action of $\mathcal{S}$ on $ \mathcal{A}_{(ii)} \cap \mathbb{W}^{\mathpzc{2},\mathpzc{2}}(\mathbb{R}^3)$ we need to calculate the Laplacian derivative and time derivative of $\Psi(t,\mathbf{x})$ from (\ref{defPSI}), (\ref{gradPhi}), (\ref{Phixx}) and (\ref{Phixy}), i.e.,
 {\small
\begin{eqnarray}
\nabla^2 \Psi (t,\mathbf{x}) & =& \negthickspace \negthickspace \left(  (\mathpzc{q}^2 - \frac{2\mathpzc{q}}{r} )\mathcal{P} +(  \nabla^2 \mathcal{P}   -2 \mathpzc{q} \hat{\mathbf{r}} \cdot \vec{\nabla} \mathcal{P})  \right) \exp{(\mathpzc{i} \varpi_\mathpzc{o} t-\mathpzc{q}r)} , \label{FOXEQ} \\
\partial_t \Psi (t,\mathbf{x}) &=& ( \mathpzc{i} \varpi_\mathpzc{o} \mathcal{P}+\partial_t \mathcal{P} ) \exp{(\mathpzc{i} \varpi_\mathpzc{o} t-\mathpzc{q} r)}. \label{FOXDT}
\end{eqnarray} }
 Theorem \ref{Sobolema3} can be generalized for the larger class of quasi-polynomials of $\varphi(t,\mathbf{x})$ \textit{and} $r=|\mathbf{x}|$, but the definition of $\mathcal{A}_{(ii)}$ given above is enough for our purposes.  Since we are not interested in nonlinear terms, we henceforth restrict to the linear case $\mathcal{P}(\varphi(t,\mathbf{x}) )=\varphi(t,\mathbf{x})$. Substituting $\mathcal{P}(\varphi(t,\mathbf{x}))=\varphi(t,\mathbf{x})$ and Eqs.   (\ref{FOXEQ}) and (\ref{FOXDT}) into Eq. (\ref{APDE}), and further using (\ref{regular2R}), yields
{\small
\begin{eqnarray}
 \mathcal{S}(\Psi) \negthickspace \negthickspace \negthickspace &=&  \mathpzc{r}_{\mathsmaller{B} } \bigg(- \delta \mathpzc{a}_{ \mathpzc{p} \upmu } +  \varLambda  \varphi 
+\mathpzc{i}\mathpzc{k}_{\; \mathsmaller{B}} (\partial_t \varphi -\mathpzc{i}\mathpzc{k}_{\; \mathpzc{q}}\varphi)- \mathpzc{q}(\partial_t \varphi^\dagger   -\mathpzc{i}\mathpzc{k}_{\; \mathpzc{q}}  \varphi^\dagger)  \bigg)\exp{(\mathpzc{i} \varpi_\mathpzc{o} t-\mathpzc{q} r)} \notag \\
 &&+ \frac{ \mathpzc{r}_{\mathsmaller{B} } }{r} \bigg(\delta \mathpzc{v}_{ \mathpzc{p} \upmu}+\mathpzc{k}_{\; \lambda}  \varphi +   ( \partial_t \varphi^\dagger   -\mathpzc{i}\mathpzc{k}_{\; \mathsmaller{D}}  \varphi^\dagger)  \bigg)\exp{(\mathpzc{i} \varpi_\mathpzc{o} t-\mathpzc{q} r)} \notag \\ 
&&+\mathpzc{R}_{\, \,\mathsmaller{ID}}(t,\mathbf{x}),  \label{APDE2}
\end{eqnarray} }
where we have defined some useful combinations of our real parameters by
{\small
\begin{eqnarray}
\varLambda & \equiv&\frac{ (\mathpzc{q}^2 +\mathpzc{k}^2_{\; \mathpzc{q}} )}{2}  -\frac{(\mathpzc{h}_{\mathsmaller{B}} \varpi_\mathpzc{o} +\mathpzc{h}_\mathsmaller{B}\mathpzc{k}_{\;\mathsmaller{D}} )}{ \mathpzc{r}_\mathsmaller{B}}, \label{defLamb} \\
\mathpzc{k}_{\;\mathsmaller{B}}& \equiv & \frac{\mathpzc{h}_\mathsmaller{B}}{\mathpzc{r}_\mathsmaller{B} } -\mathpzc{k}_{\; \mathpzc{q}}.  \label{defKB}
\end{eqnarray}}
In Eq. (\ref{APDE2}), the composite spin-orbit reminder {\small $ \mathpzc{R}_{\,\, \mathsmaller{ID}}(t,\mathbf{x}) $} is 
{\small
\begin{eqnarray}
 \mathpzc{R}_{\,\,\mathsmaller{ID}}(t,\mathbf{x}) & \equiv &  \frac{\alpha}{2} \mathpzc{r}_\mathsmaller{B}\bigg(   \mathcal{R}_{\mathpzc{p} \upmu}^+(t,\mathbf{x}) -2\mathpzc{i}\mathpzc{k}_{\; \mathpzc{q}} \mathcal{R}_{\; \Delta}^+(t,\mathbf{x})      \bigg)  \exp{(\mathpzc{i} \varpi_\mathpzc{o} t-\mathpzc{i} \varpi_\mathsmaller{q}t-\mathpzc{i} \mathpzc{k}_{\; \mathpzc{q}} r-\mathpzc{q}r )}   \notag \\ &&+ \frac{\beta}{2}  \mathpzc{r}_\mathsmaller{B}\bigg(   \mathcal{R}_{\mathpzc{p} \upmu}^- (t,\mathbf{x})  +  2\mathpzc{i}\mathpzc{k}_{\; \mathpzc{q}}  \mathcal{R}_{\; \Delta}^-(t,\mathbf{x})     \bigg)  \exp{(\mathpzc{i} \varpi_\mathpzc{o} t-\mathpzc{i} \varpi_\mathsmaller{q}t +\mathpzc{i} \mathpzc{k}_{\; \mathpzc{q}}r-\mathpzc{q}r )}. \label{remindID} 
\end{eqnarray}
}
According to (\ref{errovarphiboth}) and (\ref{resto_K}), the reminder (\ref{remindID}) belongs to $ \{  \mathbb{L}^2(\mathbb{R}^3) \cap O(\frac{\Delta_\mathpzc{s}^\pm}{r^2}) \}$. The $\delta \mathpzc{v}_{ \mathpzc{p} \upmu}(t,\mathbf{x})$ and $\delta \mathpzc{a}_{ \mathpzc{p} \upmu}(t,\mathbf{x})$ defined by (\ref{epsonV}) and (\ref{epsonA}) are bounded because the velocities and accelerations along the musical orbit are bounded. According to theorem  \ref{CONTI_BOUNDS} and Eqs. (\ref{rotating}) and (\ref{dagger}), the oscillatory functions $\varphi(t,\mathbf{x})$, $\partial_t \varphi(t,\mathbf{x})$, $\varphi^\dagger(t,\mathbf{x})$, $\partial_t \varphi^\dagger(t,\mathbf{x})$ belong to $ \mathbb{L}^\infty (\mathbb{R} \times \mathbb{R}^3)$, and therefore the whole expression inside the larger parenthesis on the first line of the right-hand side of (\ref{APDE2}) belongs to $\mathbb{L}^2_{loc}(\mathbb{R}^3)$. The term with the $\frac{1}{r}$ singularity inside the larger parenthesis of the \textit{second} line of the right-hand side of (\ref{APDE2}) belongs to $\mathbb{L}^2_{loc}(\mathbb{R}^3)$ as well.  After the multiplication by {\small $\exp{(\mathpzc{i} \varpi_\mathpzc{o} t-\mathpzc{q}r}) $}, the right-hand side of (\ref{APDE2}) belongs to $ \mathbb{L}^2 (\mathbb{R}^3)$ for any $\mathpzc{q}>0$.  

\subsection{Ordering, Fredholm-Schroedinger PDE, and spin-orbit reminder}
\label{Fredholm-Schroe}
Here we \textit{postulate} one can adjust the orbit to satisfy a matched asymptotic ordering for the terms of the right-hand side of identity (\ref{APDE2}), which can also be interpreted as linear bridging approximations to $\delta \mathpzc{a}_{ \mathpzc{p} \upmu} $ and $\delta \mathpzc{v}_{ \mathpzc{p} \upmu}$, as follows.
\begin{enumerate}
\item  The first line of the right-hand side of (\ref{APDE2}) is bounded by theorem \ref{CONTI_BOUNDS}
and because the accelerations are bounded by the equations of motion (\ref{Hans_Daniel}) if the velocities are bounded and the orbit is non-collisional. We postulate that one can adjust the orbit in order for the first line on the right-hand side of (\ref{APDE2}) to vanish for any $(t,\mathbf{x})$, yielding 
{\small
\begin{equation}
\mathpzc{C}_\mathsmaller{0}(t,\mathbf{x})=- \delta \mathpzc{a}_{ \mathpzc{p} \upmu} +\varLambda \varphi +\mathpzc{i} \mathpzc{k}_{\; \mathsmaller{B}} (\partial_t \varphi  -\mathpzc{i} \mathpzc{k}_{\; \mathsmaller{D}} \varphi )  -\mathpzc{q} (\partial_t \varphi^\dagger  -\mathpzc{i} \mathpzc{k}_{\; \mathsmaller{D}} \varphi^\dagger )=0. \label{zeroC0}
\end{equation}}
\item The second line on the right-hand side of (\ref{APDE2}) is a sum of bounded terms multiplied by $\frac{1}{r}$. We postulate one can adjust the orbit in order for the second line on the right-hand side of (\ref{APDE2}) to vanish for any $(t,\mathbf{x})$, yielding
{\small
\begin{eqnarray}
\mathpzc{C}_{\mathsmaller{1}}(t,\mathbf{x})=\delta \mathpzc{v}_{ \mathpzc{p} \upmu}+ \mathpzc{k}_{\; \lambda} \varphi + \partial_t \varphi^\dagger  -\mathpzc{i} \mathpzc{k}_{\; \mathsmaller{D}}  \varphi^\dagger   =0. \label{zeroC1}
\end{eqnarray}}
\end{enumerate}
Notice that \textit{if} (\ref{zeroC0}) and (\ref{zeroC1}) hold, then Eqs. (\ref{regular2R}) and (\ref{APDE2}) become \textit{linear PDEs} by disregarding the respective spin-orbit reminders. If the orbit satisfies (\ref{zeroC0}) and (\ref{zeroC1}), then we can re-write (\ref{APDE2}) as an effective linear Schroedinger equation in $\mathbb{W}^{\mathpzc{2},\mathpzc{2}}(\mathbb{R}^3)$ having a forcing term in $\{  \mathbb{L}^2(\mathbb{R}^3) \cap O(\frac{\Delta_\mathpzc{s}^\pm}{r^2}) \}  $, i.e.,
{\small
\begin{eqnarray}
\frac{\mathpzc{r}_\mathsmaller{B}}{2} \nabla^2 \Psi + \frac{1}{r} \Psi  +\mathpzc{i} \mathpzc{h}_\mathsmaller{B}\partial_t \Psi \negthickspace &=&\negthickspace  \mathpzc{R}_{\; \mathsmaller{ID}}(t,\mathbf{x}), \label{Fredholm-Schroedinger} 
 \end{eqnarray}
 } 
 where $ \mathpzc{R}_{\; \mathsmaller{ID}}(t,\mathbf{x}) \in  \{  \mathbb{L}^2(\mathbb{R}^3) \cap O(\frac{\Delta_\mathpzc{s}^\pm}{r^2}) \} $ is the spin-orbit reminder. Henceforth (\ref{Fredholm-Schroedinger}) is called the Fredholm-Schroedinger PDE problem.

\section{Applications of the Fredholm-Schroedinger PDE problem}
 \label{applications_Chemical_principle}
 
 Here we discuss applications of the Fredholm-Schroedinger PDE problem (\ref{Fredholm-Schroedinger}) motivated by the importance of the Schroedinger equation in physics and its open relations with electrodynamics\cite{Mehra,FeyNobel}. The idea is to formulate a problem involving some property compatible with two-body orbits of musical type.   Having chosen the property of interest, we can optimize the coefficients $ \mathpzc{r}_\mathsmaller{B} $ and $ \mathpzc{h}_\mathsmaller{B}$ along with $\mathpzc{k}_{\;\mathpzc{q}} $, $\varpi_\mathpzc{o}$ and $\varpi_{\mathpzc{q}}$ in order to satisfy the breaking-point boundary-layers while minimizing the norm of the forcing term of (\ref{Fredholm-Schroedinger}) in the chosen family of orbits, as discussed in \S  \ref{applications_Chemical_principle}-\ref{generalized} below.
 \par

 \subsection{Chemical Principle criterion}
 \label{definition_of_Chemical}
 
Reference\cite{cinderela} introduced the Chemical Principle criterion by using a resonance condition to select musical-like orbits with vanishing far-fields and a boundary-layer perturbation theory to calculate atomic spectra with a quantitative and a qualitative agreement. In the following, we formulate an educated version of the resonance condition tested in \cite{cinderela}. The asymptotic vanishing of the electric and magnetic far-fields (\ref{Efar_CPE}) and (\ref{Bfar_CPB}) of a bounded orbit requires
 {\small
 \begin{eqnarray}
 \lim_{| \mathbf{x} | \rightarrow \infty} |  \hat{\mathbf{r}}\times (\mathbf{J}^\pm_\mathpzc{p}-\mathbf{J}^\pm_\mathpzc{e}) | =0, \label{Chemical_DB}
 \end{eqnarray}} 
where the $ \mathbf{J}_\mathpzc{j}^\pm (t,\mathbf{x})$ are defined by (\ref{defAj}) for $\mathpzc{j} \in\{  \mathpzc{p},\mathpzc{e} \}$.  Equation (\ref{Chemical_DB}) is henceforth called the Chemical Principle condition.

 \subsection{Quasi-semiflow condition} 
\label{quasi-semiflow}
 At distant points way from the bounded orbit and along a direction $\hat{\mathbf{r}} \equiv \mathbf{x}/|\mathbf{x}|$, we can eliminate $t$ and $r=|\mathbf{x}|$ from Eqs. (\ref{light-cone_map}) and (\ref{defi}) in favour of $t_\mathpzc{e}$ and $t_\mathpzc{p}$. In the limit when $\max \{ |\mathbf{x}_\mathpzc{p}(t)| \}  \rightarrow 0$ we obtain
\begin{eqnarray}
t_\mathpzc{p}=t_\mathpzc{e} \pm \hat{\mathbf{r}} \cdot \mathbf{x}_\mathpzc{e}(t_\mathpzc{e}) \label{both_signs}.
\end{eqnarray}
Using Eq. (\ref{both_signs}) and definition (\ref{defAj}) for $\mathpzc{j}\in \{ \mathpzc{p},\mathpzc{e} \}$, we can cast condition (\ref{Chemical_DB}) for both signs in a single formula, i.e.,
\begin{eqnarray}
\frac{ \mathbf{a}_\mathpzc{p}(t_\mathpzc{p})  - \hat{\mathbf{n}}^{  \mathsmaller{\infty} } \times (\mathbf{v}_\mathpzc{p}(t_\mathpzc{p}) \times \mathbf{a}_\mathpzc{p}(t_\mathpzc{p})) }{ (1+\hat{\mathbf{n}}^{ \mathsmaller{ \infty} }    \cdot \mathbf{v}_\mathpzc{p}(t_\mathpzc{p}))^3     }  &=& \frac{ \mathbf{a}_\mathpzc{e}(t_\mathpzc{e})  - \hat{\mathbf{n}}^{  \mathsmaller{\infty} } \times (\mathbf{v}_\mathpzc{e}(t_\mathpzc{e}) \times \mathbf{a}_\mathpzc{e}(t_\mathpzc{e})) }{  (1+\hat{\mathbf{n}}^{ \mathsmaller{ \infty} }    \cdot \mathbf{v}_\mathpzc{e}(t_\mathpzc{e}) )^3      } , \label{PCPF_semi-flow}
\end{eqnarray}
with
\begin{eqnarray}
t_\mathpzc{p}& \equiv &t_\mathpzc{e} + \hat{\mathbf{n}}^{ \mathsmaller{ \infty} } \cdot \mathbf{x}_\mathpzc{e}(t_\mathpzc{e}) \label{asy_one_sign},
\end{eqnarray}
where $\hat{\mathbf{n}}^{ \mathsmaller{ \infty} }$  is an arbitrary unit vector in $\mathbb{R}^3$. Equation (\ref{PCPF_semi-flow}) is an equivalent version of the Chemical Principle criterion (\ref{Chemical_DB}) along bounded orbits, henceforth called the \textit{quasi-semiflow condition}.

 \subsection{Some tested consequences of the Chemical Principle criterion}
 \label{tested_consequences}

 In the following we start from (\ref{PCPF_semi-flow}) to derive the resonance condition that was used as a criterion in \cite{cinderela} in order to calculate atomic spectra. The linearized modes about circular orbits used in \cite{cinderela} are explosive and must be combined in pairs along $C^2$ segments that terminate in breaking points inside \textit{boundary-layers} where the electronic velocity goes near the speed of light and bounces discontinuously (see Figures 5 and 6 of \cite{cinderela}). The derivation of the resonance condition used in \cite{cinderela} assumes a large $M_\mathpzc{p}$ and sets $|\mathbf{v}_\mathpzc{p}| \simeq 0$ and $|\mathbf{v}_\mathpzc{e}| \simeq 1$ into the left-hand side of (\ref{PCPF_semi-flow}). Equation (\ref{PCPF_semi-flow}) then determines $\mathbf{a}_\mathpzc{p}(t_\mathpzc{p})$ \textit{and} implies that the nonlinear cross product on the right-hand side of (\ref{PCPF_semi-flow}) should cancel the electronic acceleration in order for $\mathbf{a}_\mathpzc{p}(t_\mathpzc{p})$ to be small when the electronic denominator becomes singular on the right-hand side of (\ref{PCPF_semi-flow}).  Adopting the notation defined in Eqs. (32) and (33) of Ref. \cite{ cinderela}, we express the orbital period as $\Omega \equiv 2\pi/T \equiv \theta/r_b$ and the (fast) frequencies of the mutually perpendicular resonant modes used in Ref. \cite{cinderela} by $(n\pi +\epsilon_z)/r_b $ and $(n\pi+\epsilon_{xy})/r_b$.  Equation (71) of Ref. \cite{cinderela} is obtained by averaging the term with the cross product on the right-hand side of (\ref{PCPF_semi-flow}), yielding $\theta=\epsilon_z-\epsilon_{xy}$, which is the condition solved with a Newton method in Ref. \cite{cinderela} to calculate spectroscopic lines for atoms with various values of $M_\mathpzc{p}$. In Table 1 (see pg. 179 of \cite{cinderela}) and in Tables  2  and 3  (see pg. 180  of Ref. \cite{cinderela}) we calculated the emission lines of hydrogen and muonium, respectively, and in Table 4 (see pg. 181 of  Ref.\cite{cinderela}) we calculated the emission lines of the positronium atom.\subsection{Boundary-layer theory and generalized Chemical Principle conditions}
\label{generalized}
Here we work in the opposite direction and derive a condition of the Chemical Principle type \textit{from} the linearizing conditions (\ref{zeroC0}) and (\ref{zeroC1}). Starting from the alternative family of musical orbits satisfying (\ref{zeroC0}) and (\ref{zeroC1}) we have 
{\small
 \begin{eqnarray}
\delta \mathpzc{v}_{ \mathpzc{p} \upmu} &=& -\mathpzc{k}_{\; \lambda}  \varphi - (\partial_t \varphi^\dagger  -\mathpzc{i} \mathpzc{k}_{\; \mathsmaller{D}}  \varphi^\dagger) , \label{linearVPHI} \\
 \delta \mathpzc{a}_{ \mathpzc{p} \upmu } &=&\varLambda \varphi + \mathpzc{i} \mathpzc{k}_\mathsmaller{B}(\partial_t \varphi  -\mathpzc{i} \mathpzc{k}_{\; \mathsmaller{D}}  \varphi)- \mathpzc{q}(\partial_t \varphi^\dagger  -\mathpzc{i} \mathpzc{k}_{\; \mathsmaller{D}}  \varphi^\dagger). \label{linearAPHI} 
 \end{eqnarray} }
Using (\ref{Qone_p}), (\ref{Qone_mu}), we find by inspection that the $\frac{1}{r}$ term on the right-hand side of (\ref{regular2R}) vanishes at breaking points and therefore theorem \ref{CONTI_BOUNDS} suggests a linear approximation in terms of the natural oscillatory functions and their derivatives. We henceforth restrict to the cases where either $(\alpha,\beta)=(1,0)$ or $(\alpha,\beta)=(0,1)$, in order to cancel the compensating exponential from both sides of (\ref{linearVPHI}) after expressing it in terms of $\Delta^\pm_\mathpzc{s}$ and $\partial_t \Delta^\pm_\mathpzc{s}$ using (\ref{rotating}), (\ref{dagger}) and (\ref{epsonV}). After cancelling the oscillatory exponential factor from both sides of (\ref{linearVPHI}), we find by inspection that we must take $\mathpzc{k}_{\; \mathpzc{q}}=0$ in order for the right-hand side of (\ref{linearAPHI}) to be real as required by its left-hand side. Using (\ref{rotating}) and (\ref{defKDBROG}) with $\mathpzc{k}_{\; \mathpzc{q}}=0$ to evaluate the right-hand side of (\ref{linearVPHI}), and evaluating the left-hand side of (\ref{linearVPHI}) using (\ref{epsonV}) when $(\alpha,\beta)=(1,0)$ or $(\alpha,\beta)=(0,1)$ yields
{\small
\begin{eqnarray}
\mathpzc{Q}_{\, \, \mathpzc{p}}^\pm -\mathpzc{Q}_{\, \, \upmu}^\pm  &=& - \mathpzc{k}_{\; \lambda} (\phi_\mathpzc{p}^\pm -\phi_\upmu^\pm)  , \label{balancedVPHI} \end{eqnarray}
}
where $\mathpzc{Q}_{\, \, \mathpzc{p}}^\pm $ and $\mathpzc{Q}_{\, \, \upmu}^\pm $ are defined by (\ref{Qone_p}) and (\ref{Qone_mu}), and we have used (\ref{NATURAL_OSCILLATORY}). According to Lemma \ref{musicallema}, theorem \ref{CONTI_BOUNDS} and definition (\ref{epsonV}), both sides of (\ref{balancedVPHI}) vanish at breaking points. The left-hand side of (\ref{balancedVPHI}) vanishes at breaking points by (\ref{epsonV}) and properties (P2) and (P3), while the right-hand side of (\ref{balancedVPHI}) vanishes by theorem \ref{CONTI_BOUNDS},  thus making (\ref{balancedVPHI}) a bridging linearization, as discussed in \S \ref{PDE} -\ref{estima_e_classifica}, and defining $\mathpzc{k}_{\; \lambda} \equiv \frac{1}{\mathpzc{r}_\mathsmaller{B} } -\mathpzc{q} > 0$ as a local derivative at breaking points. 
\par
The same bridging linearization is posed again by Eq. (\ref{linearAPHI}), which right-hand side vanishes at breaking points according to theorem \ref{CONTI_BOUNDS}. Therefore, in view of (\ref{epsonA}), Lemma \ref{musicallema} and theorem \ref{CONTI_BOUNDS}, the left-hand side of (\ref{linearAPHI}) must vanish at breaking points as well, thus
 requiring the \textit{fourth} musical property ($P4$) that the protonic accelerations $\mathbf{a}_\mathpzc{p}^\pm(t,\mathbf{x})$ jump at breaking points in order to match the synchronized jump of the image accelerations $\mbox{\Large $a$}_\upmu^\pm (t,\mathbf{x}) $. Boundary-layer property ($P4$) is important only for the compatibility of the musical orbit with the equations of motion (\ref{Hans_Daniel}) and it is not used here.  
\par
Conditions (\ref{linearVPHI}) and (\ref{linearAPHI}) transform (\ref{regular2R}) into a linear equation and, by using (\ref{epsonV}) and (\ref{epsonA}), we find that (\ref{linearVPHI}) and (\ref{linearAPHI}) are of the same form of the Chemical Principle condition (\ref{PCPF_semi-flow}), as follows.  Both (\ref{PCPF_semi-flow}) and (\ref{linearAPHI}) are of neutral-delay type involving accelerations divided by cubed denominators and evaluated in the past and in the future. The difference is that while (\ref{PCPF_semi-flow}) involves the delayed protonic accelerations and the delayed electronic accelerations, Eq. (\ref{linearAPHI}) involves the delayed protonic accelerations and the delayed accelerations of the \textit{image} trajectory (\ref{defimageXmu}). Recalling that the image trajectory (\ref{defimageXmu}) is made of a (slowly varying) $C^2$ guiding function $-\mathtt{G}_\mathsmaller{B}(t)$ \textit{plus} the electronic trajectory $\mathbf{x}_\mathpzc{e}(t) $, Eqs. (\ref{zeroC0}) and (\ref{zeroC1}) relate the same accelerations of the far-field compositions (\ref{Efar_CPE}) and (\ref{Bfar_CPB}). In the special case when $\tau=0$ \textit{and} $\upmu= 1$, (\ref{PCPF_semi-flow}) and (\ref{linearAPHI}) involve the \textit{same} delayed accelerations combined with the \textit{same} linear coefficient. Since (\ref{PCPF_semi-flow}) is equivalent to the Chemical Principle condition (\ref{Chemical_DB}), we postulate (\ref{zeroC0}) and (\ref{zeroC1}) as the \textit{generalized Chemical Principle conditions}.

\subsection{Linearization of the natural PDE and parameter estimates}
\label{linearization_estimate}

In the case $\mathpzc{q}=0$ and $\mathpzc{k}_{\; \mathpzc{q}}=0$, the Schroedinger equation for the hydrogen atom follows from (\ref{SOBRE_NATURAL}) by substituting (\ref{zeroC1}) and (\ref{zeroC0}) into (\ref{SOBRE_NATURAL}), multiplying by $\frac{1}{2}\mathpzc{r}_\mathsmaller{B}$ and ignoring the spin-orbit reminder, yielding
{\small
\begin{eqnarray}
\frac{\mathpzc{r}_\mathsmaller{B}}{2} \nabla^2 \varphi(t,\mathbf{x})
& =& -\mathpzc{i}\mathpzc{h}_\mathsmaller{B} \partial_t \varphi + \bigg( \frac{1}{2}\mathpzc{h}_\mathsmaller{B}\varpi_\mathpzc{o} -\frac{1}{r} \bigg) \varphi, \label{eigenSchroedinger}
\end{eqnarray} }
where we have used (\ref{defLamb}) and (\ref{defKB}). The Schroedinger equation (\ref{eigenSchroedinger}) is known to possess solutions in $\mathbb{W}^{\mathpzc{2},\mathpzc{2}}(\mathbb{R}^3)$. Recalling that theorem \ref{Sobolema3} was proved for $\mathpzc{q}>0$ only, we notice that in cases when $\varphi(t,\mathbf{x})$ itself belongs to $\mathbb{W}^{\mathpzc{2},\mathpzc{2}}(\mathbb{R}^3)$, definition (\ref{defPSI}) can be extended to $\mathpzc{q}=0$ because the decreasing exponential $\exp(-\mathpzc{q}r) $ is no longer necessary in order for (\ref{defPSI}) to be in $\mathbb{W}^{\mathpzc{2},\mathpzc{2}}(\mathbb{R}^3)$. 
\par
The use of L'H\^{o}pital's rule to perform the limit from either the left-hand side or from the right-hand side of breaking points using Eq. (\ref{balancedVPHI}) involves a ratio between acceleration differences and velocity differences.  Since the velocities are bounded and the orbit is non-collisional, the equations of motion (\ref{Hans_Daniel}) define \textit{bounded} accelerations and Eq. (\ref{balancedVPHI}) defines $\mathpzc{r}_\mathsmaller{B}  \equiv (\frac{1}{\mathpzc{q}+\mathpzc{k}_{\; \lambda}})$ directly from the orbit's boundary-layer \cite{cinderela,Mallet-Paret_bound-layer}, as estimated next. Disregarding the acceleration of the guiding trajectory and \textit{assuming} that the spiky orbits illustrated in Figures 5 and 6 of Ref. \cite{cinderela} are approximations for a musical orbit satisfying the Chemical Principle, an estimate for the Bohr radius parameter $\mathpzc{r}_{\mathsmaller{B}}$ is obtained using Eq. (33) of Ref. \cite{cinderela} ($ r_{bk}= \frac{1}{\theta_k^2} \equiv \ell_z^2$), as follows. We postulate a relation between the orbital radii $r_{bk}$ and the eigenvalues $E_k$ of the Schroedinger equation  (\ref{eigenSchroedinger}) in the form of $r_{bk}  \simeq -\frac{1}{2E_k} $, in order to fit the data in the first and fourth columns of Table 1 of Ref. \cite{cinderela} with a one-parameter formula for the orbital radii, i.e., we apply a linear regression with $\ell_z^2=\mathpzc{r}_\mathsmaller{B} k^2$ for $ k=1,...,12 $. The linear regression yields a Bohr radius parameter $\mathpzc{r}_{\; \mathsmaller{B}}=(147.04)^2$ in our unit system. The accepted value today is $\mathpzc{r}_{\; \mathsmaller{B}}=(137.036)^2$ in our unit system. The approximation  of the resonant orbits of \cite{cinderela} by musical orbits should be studied further. 

\section{Summary, discussions, conjectures, and conclusion }
\label{Sumasection}
\subsection{Summary of results}
\label{sumario}
\begin{itemize}
\item We have extended the light-cone maps outside the two-body orbits to yield delay functions of $\mathbb{R} \times \mathbb{R}^3$. The existence of the delay functions $\phi_\mathpzc{e}^\pm(t,\mathbf{x})$, $\phi^\pm_\mathpzc{p}(t,\mathbf{x})$ and $\phi_\upmu^\pm(t,\mathbf{x})$ is  granted by theorem \ref{lema1}.
\item The functions $\phi_\mathpzc{e}^\pm(t,\mathbf{x})$ and $\phi^\pm_\mathpzc{p}(t,\mathbf{x})$ of theorem \ref{lema1} provide the state-dependent delays $\phi_\mathpzc{e}^\pm(t,\mathbf{x}_\mathpzc{3}(t))$ and $\phi^\pm_\mathpzc{p}(t,\mathbf{x}_\mathpzc{3}(t))$ necessary to formulate the variational three-body problem for a trajectory $\mathbf{x}_\mathpzc{3}(t) \in \mathbb{R}^3$, as explained in \S \ref{Erdman}-\ref{manyB-VE}.

\item We have introduced the musical properties at breaking points in order to construct a synchronization function in $ \mathbb{W}^{\mathpzc{2},\mathpzc{2}}(\mathbb{R}^3)$ by theorems \ref{Sobolema} and \ref{Sobolema3}. The prototype and candidate to be a musical orbit is the orbit with spiky boundary-layers illustrated in Figures 5 and 6 of Ref. \cite{cinderela}.
\item We have associated the musical orbit with a PDE identity in $ \mathbb{W}^{\mathpzc{2},\mathpzc{2}}(\mathbb{R}^3)$ and having a forcing term belonging to $\mathbb{L}^2(\mathbb{R}^3)$, i.e., Eq. (\ref{APDE2}). 
 \item We formulated the Chemical Principle criterion using the (restricted) three-body problem to define a \textit{defensive} stability of a two-body orbit against crashing into a third charge. The Chemical Principle criterion is important to physics because it predicts hydrogen-like atomic spectra successfully \cite{cinderela}. 
 \item We worked backward from the generalized conditions (\ref{zeroC0}) and (\ref{zeroC1}) to derive the generalized Chemical Principle conditions (\ref{linearVPHI}) and (\ref{linearAPHI}), which include conditions of the same type of (\ref{PCPF_semi-flow}) as a special case.
\end{itemize}

\subsection{Discussions, conjectures and conclusion}
\label{discussions}
\begin{itemize}
\item The original motivation for the name \textit{musical} orbit was to nominate the synchronization of lemma \ref{musicallema} between trajectory points at times where the two far-field contributions have their largest magnitudes (i.e., inside the boundary-layers). A synchronization near the breaking points allows the far-fields to vanish by destructive interference and satisfy (\ref{Chemical_DB}) at least near the breaking points.  The orbit with spiky boundary-layers illustrated in Figures 5 and 6 of Ref. \cite{cinderela} was our prototype for a musical orbit.
\item The Chemical Principle function (\ref{defPSI}) is the difference between the protonic time (\ref{light-cone_map}) and the image-trajectory time (\ref{synctime}), all multiplied by a complex phase, as seen by a third charge at $(t,\mathbf{x})$. For example, \textit{if} (\ref{linearVPHI}) and (\ref{linearAPHI}) hold with $\mathpzc{q}=0$ and either $(\alpha,\beta)=(1,0)$ or $(\alpha,\beta)=(0,1)$, then our \textit{Chemical Principle function} becomes $\Psi(t,\mathbf{x})=(t_\mathpzc{p}^\pm-t_\upmu^\pm)\exp(\mathpzc{i}\varpi_\mathpzc{o} t -\mathpzc{i}\varpi_\mathpzc{q} t  \mp \mathpzc{i}\mathpzc{k}_{\; \mathpzc{q}} r )$, while $\varphi(t,\mathbf{x})= \exp(-\mathpzc{i}\varpi_\mathpzc{o} t) \Psi(t,\mathbf{x})$ satisfies the hydrogenous Schroedinger equation (\ref{eigenSchroedinger}). Assuming that we can approximate $\phi^\pm_\upmu(t,\mathbf{x})$ by $ \phi^\pm_\mathpzc{e}(t,\mathbf{x})$, the modulus of the Chemical function becomes $|\Psi(t,\mathbf{x})| \simeq |\phi_\mathpzc{p}^\pm-\phi_\mathpzc{e}^\pm| $ and the vanishing of $\Psi(t,\mathbf{x})$ at the boundaries is an intuitive condition for the advanced fields  (\ref{Efar_CPE}) to interfere destructively.
\item Our synchronization theory and musical properties were developed to model an interference at-a-distance of the two-body far-fields acting on a third particle at $(t,\mathbf{x})$, as used in Ref. \cite{cinderela} to calculate atomic spectra. The rationale to develop our PDE are different from the quantum mechanical use of $| \Psi(t,\mathbf{x})|^2$ as the probability distribution for the electronic coordinate. The use of \textit{our} Chemical Principle function $| \Psi(t,\mathbf{x})|^2$ as \textit{the} probability distribution for \textit{our} electronic coordinate along \textit{our} musical periodic orbit would run into domain-related difficulties at the least, as follows. Our electronic coordinate is defined only \textit{on} the (thin) musical periodic orbit, which does not contain the large domain $(t,\mathbf{x}) \in \mathbb{R} \times \mathbb{R}^3$ of $\Psi(t,\mathbf{x})$ (e.g. see the (thin) spiky periodic orbit illustrated in Figures 5 and 6 of Ref. \cite{cinderela}). The motion along the periodic orbit is a group acting along the (thin) periodic orbit, and, as far as we can imagine, there is nothing probabilistic directly related to our $| \Psi(t,\mathbf{x})|^2$ in the bigger domain $\mathbf{x} \in \mathbb{R}^3$. 
\item All is not lost as regards compatibility with a probabilistic interpretation, and we conjecture that one might be able to adjust $\mathpzc{q}$, $\mathpzc{k}_{\; \mathpzc{q}}$, $\varpi_\mathpzc{q}$ and $\varpi_\mathpzc{o}$ in order to satisfy (\ref{zeroC0}) and (\ref{zeroC1}) for \textit{different} musical orbits while keeping $\mathpzc{r}_\mathsmaller{B}$ and $\mathpzc{h}_\mathsmaller{B}$ \textit{fixed} at some optimal values. The former could yield a universal PDE \textit{and} a naturally associated averaging procedure over different solutions of the \textit{same} universal PDE \textit{in} the large domain $\mathbf{x} \in \mathbb{R}^3$. 
\item The Chemical Principle condition (\ref{PCPF_semi-flow}) predicted the orbital magnitudes and spectral lines of atomic hydrogen \cite{cinderela}, which was the hard test for (\ref{PCPF_semi-flow}). To inspect the agreement the reader should look in section 8 of Ref. \cite{cinderela} for Table 1 on pg. 179 containing a first set of calculations matching the first eleven circular lines of hydrogen. Table 3 on pg. 180  of \cite{cinderela} lists the spectroscopic lines of the muonium atom and Table 4 on pg. 181 lists the lines calculated for the positronium atom.
\item  The generalized Chemical Principle conditions (\ref{zeroC0}) and (\ref{zeroC1}) do not have a dynamical appeal, unless we could derive the physically appealing Chemical Principle condition (\ref{PCPF_semi-flow}) from (\ref{zeroC0}) and (\ref{zeroC1}). Otherwise, to be of interest to physics, at the least one should predict the atomic magnitudes of \S \ref{applications_Chemical_principle}-\ref{tested_consequences} working directly from the generalized conditions (\ref{zeroC0}) and (\ref{zeroC1}) and the extended musical properties including the boundary-layer property (P4) discussed in the theory outlined below (\ref{balancedVPHI}). 
\item The simplest example of a two-body orbit with vanishing far-fields are the helium orbits studied with the Darwin ODE in \cite{old_helium, simple}. Unlike hydrogen, the doubly-circular orbits of helium \cite{old_helium, simple} easily satisfy the analogous of (\ref{Chemical_DB}) because both electronic accelerations have the same modulus and opposite directions, thus allowing the far-fields to vanish by destructive interference within small perturbations of the Coulomb ODE problem. The linear stability analysis used in \cite{old_helium} with a heuristic resonance condition predicted several spectral lines of helium successfully.  We conjecture that an educated generalization could replace the heuristic resonance condition of Ref. \cite{old_helium} by the concept of a generalized musical orbit. The helium equations of motion are differential-delay neutral equations with six state-dependent delays, which should accept generalized musical orbits among the solutions. We further conjecture that a generalized Chemical Principle and Fredholm-Schroedinger PDE problem for the three-body problem (two electrons plus a positive charge) could provide a way to understand the exclusion principle in helium using variational electrodynamics.
\item Newton's third law limits the protonic to electronic acceleration ratio along Coulombian ODE orbits to $1/M_\mathpzc{p}$, and thus our derivation differs from Nelson's derivation of a linear Schroedinger equation from Newtonian mechanics \cite{Nelson}. Moreover, we start from the variational generalization of Wheeler-Feynman's electrodynamics \cite{JDE1,cinderela} rather than from the neighborhood of some orbit of a Hamiltonian ODE \cite{Nelson}. Our derivation is limited to the rotating singularities which appear naturally in the second derivative of the delay functions, i.e., those of type $1/|\mathbf{x}-\mathbf{x}_o (t) |$. Fortunately, the former class covers atomic physics, chemistry \textit{and} variational electrodynamics \cite{JDE1}. 
\item The failure of the Wheeler-Feynman ODE quantization program \cite{Mehra,FeyNobel} had at least two causes: (i) the misleading analogy with finite-dimensional Hamiltonian ODE's, which inconsistency was harder to detect before the no-interaction theorem \cite{no-interaction} and (ii) the theory of delay equations as infinite-dimensional problems  \cite{JDE1,Mallet-Paret1,Mallet-Paret_bound-layer,JackHale,Nicola2,Dirk} was not out yet.
\end{itemize}

 \section{Appendix: Usable expressions, definitions and limits}
\label{asymptotics}

In this Appendix, we derive some useful expressions involving derivatives of delay functions and asymptotic limits that are used throughout the paper. For brevity of notation we drop the upper indices $(\mathpzc{\pm})$ indicating the advanced and the retarded delay functions and extend the lower index to $\mathpzc{j} \in \{ \mathpzc{p},\mathpzc{e}, \upmu \}$, an economy of notation. Using (\ref{d2fidt2}) together with the $y$ and $z$ versions of Eqs. (\ref{d2fidxx}) and (\ref{d2fidxt}) to yield the second derivatives respect to $y$ and $z$ we obtain
{\small
\begin{eqnarray}
\nabla^2 \phi_\mathpzc{j} - \partial^2_t \phi_\mathpzc{j} =\frac{2|\vec{\nabla} \phi_\mathpzc{j}|}{\phi_\mathpzc{j}} \label{Klein},
\end{eqnarray}}
where $\nabla^2 \phi_\mathpzc{j} \equiv \partial^2_x \phi_\mathpzc{j} +  \partial^2_y \phi_\mathpzc{j} + \partial^2_z \phi_\mathpzc{j}$ is the Laplacian derivative of $\phi_\mathpzc{j}(t,\mathbf{x})$
and we have used (\ref{dndt}) to cancel the term
\begin{eqnarray}
n_x \partial_t n_x  +n_y \partial_t n_y    +n_z \partial_t n_z   =0.
\end{eqnarray}
Substituting (\ref{d2fidt2}) into (\ref{Klein}) yields the Laplacian derivative of $\phi_\mathpzc{j}(t,\mathbf{x})$ at points where $\mathbf{x}_\mathpzc{j}(t_\mathpzc{j})$ possesses two derivatives, i.e.,
{\small
\begin{eqnarray}
\negthickspace \negthickspace \negthickspace \negthickspace \negthickspace \negthickspace \nabla^2 \phi_\mathpzc{j}= -|\vec{\nabla} \phi_\mathpzc{j}|^3\hat{\mathbf{n}}_\mathpzc{j} \cdot \mathbf{a}_\mathpzc{j} +  \frac{|\vec{\nabla} \phi_\mathpzc{j}|^3  }{\phi_\mathpzc{j}}| \hat{\mathbf{n}}_\mathpzc{j} \times \mathbf{v}_\mathpzc{j} |^2   + \frac{2|\vec{\nabla} \phi_\mathpzc{j}|}{\phi_\mathpzc{j}}.
\label{Laplacian_derivative}
\end{eqnarray}}
where we have used the identity $|\hat{\mathbf{n}}_\mathpzc{j} \times \mathbf{v}_\mathpzc{j} |^2= \mathbf{v}_\mathpzc{j}^2- (\hat{\mathbf{n}}_\mathpzc{j} \cdot \mathbf{v}_\mathpzc{j})^2$. For example, the Laplacian derivative of the synchronization function can be calculated from (\ref{Laplacian_derivative}) substituting $\mathbf{v}_\mathpzc{j}(t_\mathpzc{j})$ by $\mathbf{V}_\upmu  \equiv d\mathbf{X}_\upmu/dt$ and $\mathbf{a}_\mathpzc{j}(t_\mathpzc{j})$ by $ \mbox{\Large $a$}_\upmu  \equiv d^2 \mathbf{X}_\upmu/dt^2$, yielding
 {\small
 \begin{eqnarray}
\negthickspace \negthickspace \negthickspace \negthickspace \negthickspace \negthickspace \nabla^2 \phi_\upmu= -|\vec{\nabla} \phi_\upmu |^3\hat{\mathbf{n}}_\upmu \cdot \mbox{\Large $a$}_\upmu +  \frac{|\vec{\nabla} \phi_\upmu|^3}{\phi_\upmu}|\hat{\mathbf{n}}_\upmu \times \mathbf{V}_\upmu|^2  +\frac{2|\vec{\nabla} \phi_\upmu|}{\phi_\upmu}\label{Laplamu}.
\end{eqnarray} }
The asymptotic expansions of the $\phi_\mathpzc{j}^\pm(t,\mathbf{x})$ and $\hat{\mathbf{n}}_\mathpzc{j}^\pm(t,\mathbf{x})$ defined by (\ref{defi}) and (\ref{nj}) depend only on $\mathbf{x}_\mathpzc{j}^\pm \equiv \mathbf{x}_\mathpzc{j}(t_\mathpzc{j}^+(t,\mathbf{x}))$, as follows
{\small
\begin{eqnarray}
\negthickspace \negthickspace \negthickspace \negthickspace \hat{\mathbf{n}}_\mathpzc{j}^{\pm} (t,\mathbf{x})& =& (1+\frac{\hat{\mathbf{r}} \cdot {\mathbf{x}^\pm_\mathpzc{j}}}{r}) \hat{\mathbf{r}}    -\frac{\mathbf{x}_\mathpzc{j}^\pm}{r}+ O(\frac{1}{r^2}), \label{expandnj} \\
\negthickspace \negthickspace \negthickspace \negthickspace \phi_\mathpzc{j}^\pm (t,\mathbf{x})& =& r- \hat{\mathbf{r}} \cdot \mathbf{x}_\mathpzc{j}^\pm + \frac{|\hat{\mathbf{r}} \times \mathbf{x}_\mathpzc{j}^\pm|^2}{2r}+O(\frac{1}{r^2}), \label{expandphij} \\
\negthickspace \negthickspace \negthickspace \negthickspace  t_\mathpzc{j}^\pm(t,\mathbf{x})&=&t \pm r \mp \hat{\mathbf{r}} \cdot \mathbf{x}_\mathpzc{j}^\pm \pm \frac{|\hat{\mathbf{r}} \times \mathbf{x}_\mathpzc{j}^\pm|^2}{2r}+O(\frac{1}{r^2}), \label{expandtj} \\ 
\negthickspace \negthickspace \negthickspace \negthickspace  \frac{1}{\phi_\mathpzc{j}^\pm(t,\mathbf{x})}& =& \frac{1}{r}(1 + \frac{ \hat{\mathbf{r}} \cdot \mathbf{x}_\mathpzc{j}^\pm}{r} )+ \frac{3(\hat{\mathbf{r}}\cdot \mathbf{x}^\pm_\mathpzc{j})^2-|\mathbf{x}^\pm_\mathpzc{j}|^2}{2r^3}+O(\frac{1}{r^4}), \label{expandinversephi} \\
\hat{\mathbf{r}} \cdot \hat{\mathbf{n}}_\mathpzc{j}^{\pm} (t,\mathbf{x}) &=& 1-\frac{|\hat{\mathbf{r}} \times \mathbf{x}_\mathpzc{j}^\pm    |^2 }{2r^2} + O( \frac{1}{r^3}). \label{expandrtimesnj}
\end{eqnarray}}
where $r \equiv |\mathbf{x}|$, $\hat{\mathbf{r}} \equiv \mathbf{x}/r$ and we have included an extra order in Eq. (\ref{expandrtimesnj}), beyond what is shown in (\ref{expandnj}). Since the non-analytic continuous function $\mathbf{x}_\mathpzc{j}(t)$ is never expanded, Eqs. (\ref{expandnj}), (\ref{expandphij}) and (\ref{expandinversephi}) hold at breaking points as well.

\section{Appendix: Miscellaneous about electrodynamics }
\label{Erdman}
\subsection{Equations of motion, Weierstrass-Erdmann conditions and denominators}
\label{Whee-Fey_Wey-Erd}
The conditions for a type-(ii) minimizer of the two-body problem of variational electrodynamics\cite{JDE1,cinderela} are
\begin{itemize}
\item a. To satisfy the Wheeler-Feynman equations of motion on the $\widehat{C}^2$ segments, which are expressible as
{\small
\begin{eqnarray}
\frac{ m_\mathpzc{i} \mathbf{a}_\mathpzc{i}(t)}{\sqrt{1-\mathbf{v}_\mathpzc{i}^2(t)}}&=&e_\mathpzc{i} [  \mathbf{E}_\mathpzc{j} - (\mathbf{v}_\mathpzc{i} \cdot \mathbf{E}_\mathpzc{j})\mathbf{v}_\mathpzc{i} +\mathbf{v}_\mathpzc{i} \times \mathbf{B}_\mathpzc{j} ]\label{Hans_Daniel}\\
&=&\frac{e_\mathpzc{i}}{2}\sum_{\pm} (1 \pm \hat{\mathbf{n}}_{\mathpzc{ji}}^\pm \cdot \mathbf{v}_\mathpzc{i} )\mathbf{E}_\mathpzc{j}^\pm -\frac{e_\mathpzc{i}}{2}\sum_{\pm} (\mathbf{v}_\mathpzc{i} \cdot \mathbf{E}_\mathpzc{j}^\pm)(\mathbf{v}_\mathpzc{i} \pm \hat{\mathbf{n}}_{\mathpzc{ji}}^\pm ),\notag \\ \label{Daniel_Chaos}
\end{eqnarray}}
where $\mathpzc{j}=3-\mathpzc{i}$ for $\mathpzc{i}=(1,2)$, $\mathbf{B}_\mathpzc{j}^\pm =\mp \hat{\mathbf{n}}^\pm_{\mathpzc{ji}} \times \mathbf{E}_\mathpzc{j}^\pm $ \cite{Jackson}  and $ \hat{\mathbf{n}}_{\mathpzc{ji}}^\pm $ is defined by (\ref{internormal}). Equation (\ref{Hans_Daniel}) is equal to equation (38) of Ref. \cite{cinderela}, equation (2.2) of Ref. \cite{Hans} and equation (23) of Ref. \cite{Daniel}. On the right-hand sides of (\ref{Hans_Daniel}) and (\ref{Daniel_Chaos}) are the fields of particle $j \equiv (3-i)$ at the position $\mathbf{x}_\mathpzc{i}(t)$, i.e.,
{\small
\begin{eqnarray}
 \mathbf{E}_\mathpzc{j}\equiv \mathbf{E}_\mathpzc{j}(t,\mathbf{x}_i(t)) &=& \frac{1}{2}\left( \mathbf{E}_\mathpzc{j}^+(t, \mathbf{x}_\mathpzc{i}(t))+\mathbf{E}_\mathpzc{j}^-(t, \mathbf{x}_\mathpzc{i}(t)) \right), \label{semiEj}\\
 \mathbf{B}_\mathpzc{j} \equiv \mathbf{B}_\mathpzc{j}(t,\mathbf{x}_i(t)) &=& \frac{1}{2}\left(  \mathbf{B}_\mathpzc{j}^+(t, \mathbf{x}_\mathpzc{i}(t))+\mathbf{B}_\mathpzc{j}^-(t, \mathbf{x}_\mathpzc{i}(t)) \right)  \label{semiBj} , \\ \notag \\ \notag 
\end{eqnarray}}
as defined by the semi-sums of Li\'{e}nard-Wiechert fields (\ref{electric_CPE}) and (\ref{magnetic_CPB}). The deviating arguments on the right-hand side of Eq. (\ref{Hans_Daniel}) involve two maps from particle $\mathpzc{i}$'s time into a time along particle $\mathpzc{j}$'s trajectory: (1) the \textit{retarded} map taking the time $t_\mathpzc{i} \equiv t$ to the time $\mathfrak{t}_\mathpzc{j}^{i-}(t)$ along trajectory $\mathpzc{j}$ when particle $\mathpzc{j}$'s trajectory intersects the boundary of the past light-cone of $(t, \mathbf{x}_\mathpzc{i}(t))$  and (2) the \textit{advanced} map taking the time $t_\mathpzc{i} \equiv t$ to the time $\mathfrak{t}_\mathpzc{j}^{i+}(t)$ along trajectory $\mathpzc{j}$ when particle $\mathpzc{j}$'s trajectory intersects the boundary of the future light-cone of $(t, \mathbf{x}_\mathpzc{i}(t))$, i.e.,
{\small
\begin{eqnarray}
t \rightarrow \mathfrak{t}_\mathpzc{j}^{\mathpzc{i-}}(t) \equiv t_\mathpzc{j}^-(t, \mathbf{x}_\mathpzc{i}(t)), \label{retard_map}\\
t \rightarrow \mathfrak{t}_\mathpzc{j}^{\mathpzc{i+}}(t) \equiv t_\mathpzc{j}^+(t, \mathbf{x}_\mathpzc{i}(t)) \label{advance_map},
\end{eqnarray}}
where the $t_\mathpzc{j}^\pm(t,\mathbf{x})$ are defined by (\ref{light-cone_map}). The Li\'{e}nard-Wiechert fields (\ref{electric_CPE}) and (\ref{magnetic_CPB}) include denominators which depend on the arbitrary direction $\hat{\mathbf{n}}_\mathpzc{j}^\pm(t, \mathbf{x})$ via the inverse of the modulus of (\ref{gradient}), i.e.,
{\small
\begin{eqnarray}
(1 \pm \hat{\mathbf{n}}_\mathpzc{j}^\pm \cdot \mathbf{v}_\mathpzc{j}^\pm ) =\frac{1}{|\vec{\nabla} \phi_j^\pm (t,\mathbf{x})|}, \label{directional}
\end{eqnarray} }
henceforth the Li\'{e}nard-Wiechert \textit{directional denominators}. We define the inter-particle normal by evaluating (\ref{nj}) at $\mathbf{x}_\mathpzc{i}(t)$, i.e., 
{\small  
 \begin{eqnarray}
\hat{\mathbf{n}}_{\mathpzc{ji}}^\pm(t) \equiv \hat{\mathbf{n}}_\mathpzc{j}^\pm (t,\mathbf{x}_\mathpzc{i}(t)) = \frac{\mathbf{x}_\mathpzc{i}(t)-\mathbf{x}_\mathpzc{j}^\pm}{ |\mathbf{x}_\mathpzc{i}(t)-\mathbf{x}_\mathpzc{j}^\pm | }, 
\label{internormal}
\end{eqnarray} }
and we define the transversal momentum of the orbital velocity by
{\small
\begin{eqnarray}
  \vec{\ell}_{\mathpzc{vi}}^\pm(t_\mathpzc{i}) &\equiv& \hat{\mathbf{n}}^\pm_{\mathpzc{ji}} \times \mathbf{v}_\mathpzc{i}(t_\mathpzc{i}). \label{LVI} 
\end{eqnarray} }
Along trajectories, the directional denominators (\ref{directional}) of the fields on the right-hand side of (\ref{Hans_Daniel}) define a function of time obtained by evaluating (\ref{directional}) at $(t,\mathbf{x})=(t,\mathbf{x}_\mathpzc{i}(t))$ and using (\ref{internormal}), yielding
{\small
\begin{eqnarray}
\mathpzc{D}_\mathpzc{ji}^{\pm}(t) \equiv 1/ |\vec{\nabla} \phi_\mathpzc{j}^\pm (t,\mathbf{x}_\mathpzc{i}(t))| =1 \pm \hat{\mathbf{n}}_{\mathpzc{ji}}^\pm \cdot \mathbf{v}_\mathpzc{j}^\pm,
\label{orbital_deno}
\end{eqnarray}}
 here called \textit{velocity denominators}. 
 \item b. At breaking points, a type-(ii) minimizer of variational electrodynamics should further satisfy the Weierstrass-Erdmann continuity conditions for the partial momenta, $\mathbf{P}_\mathpzc{i}^\ell=\mathbf{P}_\mathpzc{i}^\mathpzc{r}$, and the continuity of the partial energies \cite{JDE1}, $E_\mathpzc{i}^\ell=E_\mathpzc{i}^\mathpzc{r}$, where superscripts $\ell$ and $\mathpzc{r}$ indicate the quantity on the left-hand side and on the right-hand side of the discontinuity point, respectively, and
 {\small
\begin{eqnarray}
\mathbf{P}_\mathpzc{i} ^{ \{ \mathpzc{r}, \ell \} }&  \equiv &  \frac{m_\mathpzc{i} \mathbf{{v}}_\mathpzc{i}  }{ \sqrt{1 - \mathbf{v}_\mathpzc{i}^2}  }   
- \left.  \left( \frac{\mathbf{{v}}_\mathpzc{j}^ -  }{2r_\mathpzc{ji}^{-}\mathpzc{D}_\mathpzc{ji}^-(t) }  +
\frac{\mathbf{{v}}_\mathpzc{j }^+}{2 r_\mathpzc{ji}^{+}\mathpzc{D}_\mathpzc{ji}^+(t)} \right)  \right\vert_{\mathpzc{r}, \ell},
\label{partial_momentumi} \\
E_\mathpzc{i}^{ \{ \mathpzc{r}, \ell \} }  & \equiv & \frac {m_\mathpzc{i}} {\sqrt{1-\mathbf{v}_\mathpzc{i}^2}}  - \left. \left(  \frac{1}{2r_\mathpzc{ji}^{-} \mathpzc{D}_\mathpzc{ji}^- (t)}  +
\frac{1}{2r_\mathpzc{ji}^{+}\mathpzc{D}_\mathpzc{ji}^+(t)} \right)  \right\vert_{\mathpzc{r}, \ell}   \label{partial_energyi},
\end{eqnarray}}
where $r_\mathpzc{ji}^{\pm}$ are the (continuous) inter-particle distances 
{\small
\begin{eqnarray}
 r_\mathpzc{ji}^{\pm} &\equiv& \phi_\mathpzc{j}^\pm(t,\mathbf{x}_\mathpzc{i}(t)).\label{inter_distance}
\end{eqnarray}}
The subscripts $ \mathpzc{r}$ and $\ell $ on the right-hand side of (\ref{partial_momentumi}) and (\ref{partial_energyi}) indicate the right-hand side and the left-hand side each breaking point.
\end{itemize}

\par
\subsection{Li\'{e}nard-Wiechert vector-fields}
\label{Lienard_fields}
 The Li\'{e}nard-Wiechert vector-fields of a point charge \cite{Jackson} are
 {\small
\begin{eqnarray}
\mathbf{E}_\mathpzc{j}^\pm(t, \mathbf{x}) &\equiv &e_\mathpzc{j}(1-|\mathbf{v}_\mathpzc{j}^\pm|^2)\frac{|\vec{\nabla} \phi^\pm_\mathpzc{j}|^3}{\phi_\mathpzc{j}^2} \left(  \hat{\mathbf{n}}^\pm_\mathpzc{j}   \pm \mathbf{v}^\pm_\mathpzc{j}  \right) + e_\mathpzc{j}\frac{\hat{\mathbf{n}}^\pm_\mathpzc{j} \times (\hat{\mathbf{n}}^\pm_\mathpzc{j} \times \mathbf{J}^\pm_\mathpzc{j} )}{\phi_\mathpzc{j}}, \label{electric_CPE} \\
\mathbf{B}_\mathpzc{j}^\pm (t,\mathbf{x}) & \equiv &-e_\mathpzc{j}(1-|\mathbf{v}_\mathpzc{j}^\pm|^2)\frac{|\vec{\nabla} \phi^\pm_\mathpzc{j}|^3}{\phi_\mathpzc{j}^2} \left(  \hat{\mathbf{n}}^\pm_\mathpzc{j}   \times \mathbf{v}^\pm_\mathpzc{j}  \right)\pm e_\mathpzc{j}\frac{\hat{\mathbf{n}}^\pm_\mathpzc{j} \times \mathbf{J}^\pm_\mathpzc{j}}{\phi_\mathpzc{j}}, \label{magnetic_CPB} 
\end{eqnarray}}
where
{\small
\begin{eqnarray}
\mathbf{J}_\mathpzc{j}^\pm(t,\mathbf{x}) &\equiv &|\vec{\nabla} \phi^\pm_\mathpzc{j}|^3\bigg(\mathbf{a}_\mathpzc{j}^\pm \mp\hat{\mathbf{n}}_\mathpzc{j}^\pm \times(\mathbf{v}_\mathpzc{j}^\pm \times \mathbf{a}_\mathpzc{j}^\pm) \bigg)
 \label{defAj}.
 \end{eqnarray}}
The last terms on the right-hand sides of (\ref{electric_CPE}) and (\ref{magnetic_CPB}) are the far-fields, which define the O$(\frac{1}{r})$ asymptotic behaviour along bounded orbits, i.e., 
{\small
\begin{eqnarray}
 \negthickspace \negthickspace  \negthickspace \negthickspace  \negthickspace \negthickspace \mathbf{E}_{\mbox{\tiny$far$}}^{\pm }(t,\mathbf{x}) &  \equiv & \negthickspace \negthickspace \negthickspace  \sum_{\mathpzc{j}=\mathpzc{e},\mathpzc{p}} \frac{e_\mathpzc{j}}{\phi^\pm_\mathpzc{j}} \hat{\mathbf{n}}_\mathpzc{j}^\pm \times  \left (\hat{\mathbf{n}}_\mathpzc{j}^\pm \times \mathbf{J}_\mathpzc{j}^\pm  \right) =   \frac{1}{r}  \hat{\mathbf{r}} \times   \left (\hat{\mathbf{r}} \times   \left(  \sum_{\mathpzc{j}=\mathpzc{e},\mathpzc{p}}      e_\mathpzc{j} \mathbf{J}_\mathpzc{j}^\pm (t,\mathbf{x}) \right) \right) + O(\frac{1}{r^2}),\notag \\ \label{Efar_CPE}\\
\negthickspace \negthickspace  \negthickspace \negthickspace \negthickspace  \negthickspace \negthickspace \mathbf{B}_{\mbox{\tiny$far$}}^{\pm }(t,\mathbf{x}) &  \equiv &  \pm \sum_{\mathpzc{j}=\mathpzc{e},\mathpzc{p}} \frac{e_\mathpzc{j}}{\phi^\pm_\mathpzc{j}} \hat{\mathbf{n}}_\mathpzc{j}^\pm \times {\mathbf{J}}_\mathpzc{j}^\pm (t,\mathbf{x}) = \pm \frac{1}{r}  \hat{\mathbf{r}} \times \left(  \sum_{\mathpzc{j}=\mathpzc{e},\mathpzc{p}}    e_\mathpzc{j} {\mathbf{J}}_\mathpzc{j}^\pm (t,\mathbf{x}) \right) + O(\frac{1}{r^2}),\notag \\  \label{Bfar_CPB}
 \end{eqnarray}}
 where $\mathbf{J}_\mathpzc{j}^\pm (t,\mathbf{x}) $ is defined by (\ref{defAj}) for $\mathpzc{j} \in \{ \mathpzc{e},\mathpzc{p} \}$.  On the right-hand sides of (\ref{Efar_CPE}) and (\ref{Bfar_CPB}), we have used (\ref{expandnj}) and (\ref{expandinversephi}) in order to express the far-fields. The magnetic field (\ref{magnetic_CPB})  of a point charge is the curl of a vector potential \cite{Jackson}, i.e.,
 {\small
 \begin{eqnarray}
 \mathbf{B}_\mathpzc{j}^\pm (t,\mathbf{x}) \equiv e_\mathpzc{j} \vec{\nabla} \times \mathbf{A}^\pm_\mathpzc{j} (t,\mathbf{x}) \label{vector_potential},
 \end{eqnarray}}
 from which  we define each charge's electro-momentum vectors by 
 {\small
 \begin{eqnarray}
 \mathtt{P}^\pm_\mathpzc{j}(t,\mathbf{x})   \equiv \phi^\pm_\mathpzc{j} \mathbf{A}^\pm_\mathpzc{j} (t,\mathbf{x})     = \frac{  \mathbf{v}^\pm_\mathpzc{j}  }{ (1 \pm \hat{\mathbf{n}}_\mathpzc{j} \cdot \mathbf{v}^\pm_\mathpzc{j}) } \label{eletromomentum}.
 \end{eqnarray}}
 The divergence of the $\mathtt{P}_\mathpzc{j}^\pm$ is obtained using Eqs. (\ref{gradnv}), (\ref{defAj}) and  (\ref{eletromomentum}), yielding
{\small
\begin{eqnarray}
\vec{\nabla} \cdot \mathtt{P}^\pm_\mathpzc{j} =\mp \nabla^2 \phi_\mathpzc{j} \pm \frac{2 |\vec{\nabla} \phi^\pm_\mathpzc{j}| }{\phi^\pm_\mathpzc{j}} =\mp \partial_t^2 \phi_{\mathpzc{j}} =\pm \hat{\mathbf{n}}_\mathpzc{j} \cdot \mathbf{J}^\pm_\mathpzc{j} \mp  \frac{ |\vec{\nabla} \phi^\pm_\mathpzc{j}|^3 }{\phi^\pm_\mathpzc{j}}|\hat{\mathbf{n}}_\mathpzc{j} \times \mathbf{v}_\mathpzc{j}|^2   \label{divergentP}.
\end{eqnarray}}
Along bounded orbits, the far-fields (\ref{Efar_CPE}) and (\ref{Bfar_CPB}) depend linearly on $ \hat{\mathbf{r}}^\pm \times \mathbf{J}^\pm_\mathpzc{j} $ and are related to the curl of the electro-momentum vector (\ref{eletromomentum}) by
 {\small
\begin{eqnarray}
 \vec{\nabla} \times \mathtt{P}^\pm_\mathpzc{j} =\pm \hat{\mathbf{n}}^\pm_\mathpzc{j} \times \mathbf{J}^\pm_\mathpzc{j} +\frac{ (\mathbf{v}^2_\mathpzc{j} \pm \hat{\mathbf{n}}_\mathpzc{j}\cdot \mathbf{v}_\mathpzc{j} ) | \nabla \phi_\mathpzc{j} |^3  }{\phi_\mathpzc{j}} \hat{\mathbf{n}}^\pm_\mathpzc{j} \times \mathbf{v}^\pm_\mathpzc{j}   =\pm \hat{\mathbf{r}}^\pm \times \mathbf{J}^\pm_\mathpzc{j} + O(\frac{1}{r}).\notag \\ \label{ntimesJ_to_P}
\end{eqnarray}}

\subsection{The many-body problem of Variational Electrodynamics}
\label{manyB-VE}
The many-body problem of variational electrodynamics discussed in \cite{cinderela} has a variational problem and an equation of motion for a charge $e_\mathpzc{3}$ of mass $m_\mathpzc{3}$ acted upon by the field of all others. The equation of motion is
{\small
\begin{equation}
m_\mathpzc{3}\frac{d}{dt}\left( \frac{\mathbf{v}_\mathpzc{3}}{\sqrt{1-\mathbf{v}_\mathpzc{3}^{2}}} \right)= e_\mathpzc{3} \lbrack \mathbf{E}_\mathpzc{S}(t,\mathbf{x}_\mathpzc{3}(t))+ \mathbf{v}_\mathpzc{3
} \times \mathbf{B}_\mathpzc{S}(t,\mathbf{x}_\mathpzc{3}(t)) \rbrack \label{Lorentz},
\end{equation}}
where $\mathbf{x}_\mathpzc{3}(t)$ and $\mathbf{v}_\mathpzc{3}(t)$ are the position and velocity of the third particle\cite{JMP2009,cinderela,Jackson}. The fields on the right-hand side of Eq. (\ref{Lorentz}) are created by all other charges but $e_\mathpzc{3}$ and Eq. (\ref{Lorentz}) is supposed to hold almost-everywhere \cite{JDE1}.  Equation (\ref{Lorentz}) is actually the first of three variational conditions for a minimizer of the respective many-body problem, the other two being the Weierstrass-Erdmann corner conditions at velocity discontinuity points\cite{cinderela}, i.e., the continuity of the functions expressed by Eqs. (\ref{partial_momentumi}) and (\ref{partial_energyi}). The vector functions of time $\mathbf{E}_\mathpzc{S}(t,\mathbf{x}_\mathpzc{3}(t))$ and $\mathbf{B}_\mathpzc{S}(t,\mathbf{x}_\mathpzc{3}(t))$ on the right-hand side of (\ref{Lorentz}) are the semi-sums of vector fields (\ref{electric_CPE}) and (\ref{magnetic_CPB}) evaluated along the trajectory $\mathbf{x}_\mathpzc{3}(t)$, i.e.,
{\small
\begin{eqnarray}
 \mathbf{E}_\mathpzc{S} (t,\mathbf{x}_\mathpzc{3}(t)) & \equiv &\frac{1}{2} \sum_{j \neq 3 } \left( \mathbf{E}_\mathpzc{j}^{+}(t,\mathbf{x}_\mathpzc{3}(t)) + \mathbf{E}_\mathpzc{j}^{-}(t,\mathbf{x}_\mathpzc{3}(t)\,) \right), \label{sumE} \\
  \mathbf{B}_\mathpzc{S} (t,\mathbf{x}_\mathpzc{3}(t))& \equiv &\frac{1}{2} \sum_{j \neq 3} \left( \mathbf{B}_\mathpzc{j}^{+}(t,\mathbf{x}_\mathpzc{3}(t)) + \mathbf{B}_\mathpzc{j}^{-}(t,\mathbf{x}_\mathpzc{3}(t)\,) \right).\label{sumB}
 \end{eqnarray}}
The sign $\sum _{\mathpzc{j} \neq \mathpzc{3}}$ on the right-hand side of (\ref{sumE}) and (\ref{sumB}) is a reminder that the fields of charge $e_\mathpzc{3}$ do not contribute to its equation of motion \cite{JDE1}. For the restricted three-body problem of the Chemical Principle criterion, the right-hand sides of (\ref{sumE}) and (\ref{sumB}) include the fields of the hydrogen atom only, i.e., the electronic and the protonic fields. Notice that the equation of motion of the third charge requires the delay functions $\phi_\mathpzc{p}^\pm(t,\mathbf{x}_\mathpzc{3}(t))$ and $\phi_\mathpzc{e}^\pm(t,\mathbf{x}_\mathpzc{3}(t))$ of theorem \ref{lema1}.

\section*{Acknowledgement}
\label{FAPESP}
Grants FAPESP 2016/01948-6 and FAPESP 2016/25895-9 partially supported this work.

\end{document}